\documentclass{iopart}
\expandafter\let\csname equation*\endcsname\relax
\expandafter\let\csname endequation*\endcsname\relax

\usepackage{amsmath}
\usepackage{graphicx}
\usepackage{subfig}

\usepackage{bm}
\usepackage{float}
\usepackage{color}
\usepackage{sidecap}
\usepackage{soul}
\setstcolor{red}
\usepackage{citesort}
\usepackage{ulem}

\newcommand{\ket}[1]{{\left|  #1 \right\rangle}}
\newcommand{\bra}[1]{{\left\langle  #1 \right|}}

\begin{document}

\author{Ling-Na Wu}
\address{Max-Planck-Institut f\"ur Physik komplexer Systeme, N\"othnitzer Stra{\ss}e 38,  01187-Dresden, Germany}
\ead{lnwu@pks.mpg.de}
\author{Alexander Schnell}
\address{Max-Planck-Institut f\"ur Physik komplexer Systeme, N\"othnitzer Stra{\ss}e 38,  01187-Dresden, Germany}
\author{Giuseppe De Tomasi}
\address{Max-Planck-Institut f\"ur Physik komplexer Systeme, N\"othnitzer Stra{\ss}e 38,  01187-Dresden, Germany}
\author{Markus Heyl}
\address{Max-Planck-Institut f\"ur Physik komplexer Systeme, N\"othnitzer Stra{\ss}e 38,  01187-Dresden, Germany}
\author{Andr\'e Eckardt}
\address{Max-Planck-Institut f\"ur Physik komplexer Systeme, N\"othnitzer Stra{\ss}e 38,  01187-Dresden, Germany}
\ead{eckardt@pks.mpg.de}
\graphicspath{{figures/}}

\title{Describing many-body localized systems in thermal environments}


\begin{abstract}
In this work we formulate an efficient method for the description of many-body localized systems in weak contact with thermal
environments at temperature $T$. For this purpose we exploit the representation of the system in terms of quasi-local integrals of motion ($l$-bits) to derive a quantum master equation using Born-Markov approximations.
We show how this equation can be treated  by using
quantum-jump Monte-Carlo techniques as well as by deriving approximate kinetic equations of motion.
As an example, we consider the one-dimensional Anderson model for spinless fermions including also nearest-neighbor interactions,
which we diagonalize approximately by employing a recently proposed method valid in the limit of strong disorder and weak interactions.
Coupling the system to a global thermal bath, we study the transport between two leads with different chemical potentials at both of its ends.
We find that the temperature-dependent current is captured by an interaction-dependent version of Mott's law for variable range hopping, where transport is enhanced/lowered depending on whether the interactions are attractive or repulsive,
respectively. We interpret these results in terms of spatio-energetic correlations between the $l$-bits.
%
%
%
%
%
%
%
%
\end{abstract}
\maketitle

\section{Introduction}
Many-body localization (MBL) has emerged as a new paradigm for phase structures in interacting quantum matter protected by the underlying robust nonergodicity imposed by strong disorder~\cite{BASKO20061126,Gorn05,Altman2015MBLReview,Nandkishore2015MBLReview,ALET2018MBLReview,abanin2018ergodicity}.
This has led to the discovery of novel topological or symmetry-broken phases~\cite{Nandkishore2015MBLReview,2013PhRvB..88a4206H,2013JSMTE..09..005B,2014PhRvL.113j7204K,2014PhRvL.112u7204V,bahri2015localization}, which cannot exist in thermalizing systems, with the particularly prominent example of time crystals breaking not only spatial but also temporal symmetries~\cite{2016PhRvL.116y0401K,2016PhRvL.117i0402E,2017Natur.543..221C,2017Natur.543..217Z,2017NatPh..13..424M}.
Although many signatures of MBL have been accessed experimentally~\cite{Modugno2014PhysRevLett.113.095301,Schreiber2015Science,smith2016NP,Bordia2016PhysRevLett.116.140401,Choi2016Science,Roushan2017Science,Bordia2017PhysRevX.7.041047,Bloch2017PhysRevX.7.011034}, dissipation induced by a remaining coupling to an environment, even if weak, has turned out to have a crucial impact onto the long-time dynamics~\cite{2007PhRvB..76e2203B,Bloch2017PhysRevX.7.011034,Nandkishore2014PhysRevB.90.064203,Levi2016PhysRevLett.116.237203,Fischer2016PhysRevLett.116.160401,Medvedyeva2016PhysRevB.93.094205,Nandkishore2016AP,Everest2017PhysRevB.95.024310,Rubio2018arXiv,2018arXiv180604772L}.
Specifically, external baths with large bandwidths and delocalized excitations are expected to force MBL systems towards thermalization, which destabilizes the nonergodic properties central for the anticipated new phase structures.
However, it has remained as a challenge to theoretically describe MBL systems coupled to such environments for concrete microscopic models.
In this work we develop an efficient formalism with which we can access the steady states of MBL systems weakly coupled to thermal environments.
We apply this method to a spinless fermionic Hubbard chain with strong onsite disorder and weak interactions, where we achieve to solve for mesoscopic system sizes of the order of 100 lattice sites.
Utilizing the $l$-bit representation of MBL systems in terms of quasi-local integrals of motion, we derive a quantum master equation by means of a standard Born-Markov approximation~\cite{BreuerPetruccione}.
While deriving explicitly the $l$-bit representation is in general a demanding task, here, we make use of a recently proposed efficient approximate method controlled in the limit of strong disorder and weak interactions~\cite{Giu18}.
We show how this derived quantum master equation can be solved by means of quantum-jump Monte-Carlo simulations~\cite{Andrew2014AP}.
In order to push our description to even larger systems we use a further simplification of the quantum master equation in terms of a kinetic theory, which as we show provides also an accurate description with the advantage that it allows us to reach system sizes of the order of 100 lattice sites.
%

%
%
%
%

%
The coupling of an MBL system to an environment induces relaxation washing out information about initial conditions and thereby destabilizes the nonergodic nature of the MBL phase.
However, the dynamics in the resulting steady state can still be dominated by the localized character of the system Hamiltonian leading to slow and constrained dynamics.
A prominent example constitutes variable-range hopping, which is well-studied for Anderson localized systems of noninteracting particles~\cite{Mott1969}.
The coupling to a thermal bath induces nonvanishing transport coefficients, while the conductivity $\sigma$ remains highly suppressed at low temperatures $T$ following Mott's law,
\begin{equation}
\sigma \propto e^{- \left[{T_0}/{T}\right]^{1/(d+1)}} \, ,
\end{equation}
where $d$ denotes the spatial dimension and $T_0$ the temperature scale at which Mott's law sets in.
We take variable-range hopping as the starting point for our considerations.
For that purpose we couple our studied Hubbard chain to a bosonic thermal bath at temperature $T$.
Further, we couple the chain at both ends to fermionic reservoirs with a chemical potential difference such as to induce a current flowing through our system.
This induces a nonequilibrium current-carrying steady state, which reduces to the conventional linear response regime in the limit of vanishing chemical potential difference.
We first study a noninteracting Anderson insulator in one spatial dimension.
Starting from our microscopic model, we map out the full temperature dependence of transport.
At asymptotically low temperatures we observe that the induced current becomes temperature-independent characterized by long-range hopping particles across the full chain, occurring with an amplitude which is exponentially suppressed as a function of system size.
Upon increasing temperature beyond this finite-size dependent asymptotic regime, we recover a conductivity following Mott's law, which we take as a first indication that our approach captures the relevant physics.
At even larger temperature, transport crosses over to a simple activated regime where $\sigma \propto \exp[-T_1/T]$.
There, particles can overcome typical energy barriers imposed by the strong disorder potential on short distances as opposed to variable-range hopping which is characterized by longer-ranged tunneling processes.

As a next step, we then apply our approach to the weakly interacting case at strong disorder.
In the temperature regime, where for the Anderson system we have observed variable-range hopping, we again find that transport follows Mott's law, however, the Mott temperature $T_0$ is modified by interactions.
%
%
Depending on whether interactions are repulsive or attractive, $T_0$ increases or decreases with respect to the noninteracting Anderson insulating case.
We interpret the modifications of $T_0$ in terms of interaction-induced changes in the spatio-energetic correlations
among the local integrals of motion in the MBL system.

The paper is organized as follows. In Section 2, we present a general scheme to address MBL systems
in the presence of thermal baths. This is followed by the description of  the model under consideration  (Section 3).
In Section 4, we introduce the quantum-jump Monte-Carlo method and the kinetic theory that
we utilize to solve the master equations that  describe the dynamics of the open quantum system.
We then discuss the dependence of transport in the noninteracting system in Section~5.
In  Section~6, we address transport through the system in the presence of weak interactions.
Finally, a summary of the main results is given in Section 7 to conclude.

\section{Master equation for many-body localized systems in the presence of thermal baths}
\label{sec:Master-MBL}
 When a quantum system
is brought into weak contact with a thermal environment, the impact of the environment can be captured by quantum jumps
between different eigenstates of the system.  The rate at which such a process occurs depends on
the energy difference of the corresponding states.
For most interacting quantum systems, however, the full
spectrum is hard to obtain, so already deriving the equations of motion of the system
in the presence of a thermal environment is challenging.
We approach this problem by exploiting the fact that MBL systems posses an extensive number of integrals of motion~\cite{Huse2014PhysRevB.90.174202,ROS2015420,Imbrie2016,Rade}.
This makes them very special in the sense that they are interacting quantum systems
whose  full many-body spectrum can be accessed. We make use of this fact to
find equations of motion for MBL systems in thermal environments.

\subsection{Born-Markov master equation}
\label{sec:Born-Markov}

We are interested in the typical  setup of  open quantum systems described by the total Hamiltonian
\begin{equation}
	\hat{H}_{\mathrm{tot}} = \hat{H} + \hat{H}_{\mathrm{SB}} + \hat{H}_\mathrm{B}.
\end{equation}
Here, $\hat{H}$ denotes the Hamiltonian of the system under investigation, $\hat{H}_\mathrm{B}= \sum_\alpha \hbar \omega_\alpha \hat b^\dagger_\alpha \hat b_\alpha$
describes the bath with bosonic or fermionic annihilation operator $\hat{b}_\alpha$ for the mode with energy $\hbar\omega_\alpha$, and $\hat{H}_\mathrm{SB}$ is the system--bath
coupling operator, whose
overall strength $\gamma$ we assume to be small.
The bath is assumed to be in thermal equilibrium with temperature $T$ and chemical potential $\mu$.

In the weak-coupling limit ($\gamma \rightarrow 0$) and assuming large thermal baths,
the dynamics of the reduced density matrix $\hat\rho = \mathrm{Tr}_B\left\{\hat\rho_\mathrm{tot} \right\}$ of the system after tracing out the bath,
can be described using the Born-Markov and the rotating wave approximation. This gives rise to  a
Lindblad master equation~\cite{BreuerPetruccione}
\begin{equation}\label{rho}
{\partial _t}\hat{\rho}  =  - \frac{i}{\hbar }\left[ {{\hat H},\hat{\rho} } \right] + \sum\limits_\alpha{\left({\hat L}_\alpha \hat{\rho} {\hat L}_\alpha^\dag - \frac{1}{2} \left\{{\hat L}_\alpha^\dag {\hat L}_\alpha, \hat{\rho}\right\} \right)}\equiv - \frac{i}{\hbar }\left[ {{\hat H},\hat{\rho} } \right] + {\cal D}(\hat{\rho}),
\end{equation}
where $\{ \cdot, \cdot \}$ denotes the anti-commutator. The first term describes the unitary evolution, and the second term describes the coupling of the system to the environment.
Here the operators $\hat{L}_\alpha$ describe quantum jumps between the eignestates of the system.

Let us assume  a system with spectrum $E_k$ and eigenstates $\ket{k}$ coupled
to a phonon bath with bosonic operators $\hat{b}_\alpha$ via the system operator $\hat{v}$,
\begin{align}
\hat{H}_\mathrm{SB}=\gamma \hat{v} \otimes \sum_{\alpha} \lambda_\alpha (\hat b_\alpha + \hat b_\alpha^\dagger).
\label{eq:sb-phonon}
\end{align}
Then, the master equation takes the form
\begin{align}\label{Linbl-heatb-generl}
	\partial_t\hat{\rho} = - \frac{i}{\hbar }\left[ {{\hat H},\hat{\rho} } \right] + \sum_{k,q}  \left( \hat L_{kq} \hat \rho  \hat L_{kq}^\dagger - \frac{1}{2}\left\{ \hat L_{kq}^\dagger \hat L_{kq}, \hat \rho \right\}  \right).
\end{align}
The bath induces quantum jumps from level $q$ to $k$, mediated by the action of the jump operators
\begin{align}
\hat{L}_{kq}= \sqrt{R_{kq}}\ket{k} \bra{q} ,
\end{align}
with the jump rates
\begin{align}
	R_{kq} = \frac{2\pi \gamma^2}{\hbar} \vert \bra{k} \hat v \ket{q} \vert^2 g(E_k - E_q).
\end{align}
Here we have defined the bath correlation function
\begin{align}
	g(E) = \left\{ \begin{array}{cc}
	 J(E) n_\mathrm{B}(E, T)& \text{ for } E \geq 0\\
	  J(-E) [1+n_\mathrm{B}(-E, T)]& \text{ for } E < 0\\
	\end{array}\right.
	 \label{eq:bathcorr}
\end{align}
with the Bose distribution $n_\mathrm{B}(E, T)=(e^{E/k_\mathrm{B} T}+1)^{-1}$ and the spectral density of the bath $J(E)=\sum_{\alpha}\lambda_{\alpha}^2 \delta(E-\hbar\omega_{\alpha}) $. Note that the Born-Markov and rotating-wave approximation
rely on the assumption of a separation of timescales~\cite{BreuerPetruccione}.
For them to be valid, the timescales of the system dynamics $\tau_S \propto \vert E_k-E_q\vert^{-1}$ need to be fast when compared to the timescales of dissipation $\tau_R \propto R_{kq}^{-1}$ (we set $\hbar = k_B = 1$ hereafter).
This implies
\begin{align}\label{eq:cond-rotwave}
R_{kq} \ll |E_k - E_q|.
\end{align}

Under the dynamics of Eq.~\eqref{Linbl-heatb-generl}, the 
coherences, $\bra{k} \hat{\rho} \ket{q}$ with $k \neq q$, decay such that
the steady state reached asymptotically in the long-time limit is diagonal in the energy basis, $\hat \rho \to \hat{\rho}_\infty = \sum_k p_k \ket{k}\bra{k}$ \cite{EspositoGaspard03,HoneEtAl09,Thingna13}.
This can be seen most easily by noting that in the limit $\gamma \to 0$, the first term in Eq.~\eqref{rho} has to vanish for the steady state with $\partial_t \hat{\rho}_\infty =0$. After a diagonal state is reached, the  asymptotic dynamics  is governed by dissipation and described by the Pauli rate equation
 \begin{align}
	 \partial_t p_k = \sum_q \left(R_{kq} p_q - R_{qk} p_k\right).
	 \label{eq:Pauli}
\end{align}
Thus, the dissipative part of the master equation determines the probability distribution $p_k$ of the steady state.
When the system is coupled to a single bath of temperature $T$, the rates obey the equilibrium condition $R_{kq}/R_{qk} = e^{-(E_k-E_q)/T}$.
In this case the system approaches thermal equilibrium in the long-time limit, $p_k\propto \exp(-E_k/T)$, and the system obeys detailed balance: the terms of the sum in Eq.~\eqref{eq:Pauli}, which correspond to the net probability
currents from states $q$ to $k$, vanish individually. If the equilibrium condition is broken,  e.g.\ when the system is coupled to baths of different temperature or chemical potential, the system approaches a non-equilibrium steady state,
for which the right-hand side of Eq.~\eqref{eq:Pauli} vanishes only as a whole, so that probability currents break detailed balance.

\subsection{Many-body localized systems}
Note that in order to find the equations of motion for the system in the presence of a thermal bath, it is essential to diagonalize
the full Hamiltonian to find its eigenstates and eigenenergies.
For a quantum many-body system, this is generally difficult. Non-interacting systems however can be
treated readily \cite{DanielPRE}, since the many-body eigenstates are given by the Fock states of the single-particle Hamiltonian.
As we discuss in the following, MBL systems provide another exception,
due to an emergent form of integrability~\cite{Huse2014PhysRevB.90.174202,ROS2015420,Imbrie2016,Rade}.

We consider a system of spinless lattice fermions.
A typical example for such a fermionic system with an MBL phase is the one-dimensional Anderson model of spinless fermions with
nearest-neighbor interactions $V$, which we introduce in Section~\ref{sec:model}.
The generalization to spinful fermions or spin Hamiltonians
is straightforward.
We assume that the system Hamiltonian can be brought into the diagonal form
\begin{equation}
\hat H = \sum_k{ \varepsilon_k {\hat { n}}_k} + \frac{1}{2}\sum_{k,q}{U_{kq} {\hat { n}}_k {\hat { n}}_q }  + \frac{1}{6} \sum_{k,q,p}{U_{kqp}{\hat { n}}_k {\hat { n}}_q {\hat { n}}_p } +  \cdots,
\label{eq:MBL-hamil}
\end{equation}
where the fermionic number operators $\hat{n}_k$ describe quasi-local integrals of motion ($l$-bits) coupled by diagonal matrix elements that decay exponentially with the localization length $\xi$~\cite{Chandran15},
$U_{k_1k_2...k_n} \sim e^{- \min_{i,j} |k_i-k_j|/\xi}$ (assuming the $l$-bits to be ordered according to their spatial position).
We can directly read off the full many-particle  eigenstates, which are the  Fock states with respect to the $l$-bits
$\ket{\bf n}= \ket{ \{n_k\}}$, $n_k = 0,1$, and the corresponding  many-particle  spectrum $E_\mathbf{n}=E(\{n_k\})$.

We focus on the limit of weak interactions, where it is assumed that there exists a quasi-local
transformation (adiabatic connection) between the $l$-bits and the local annihilation operators $\hat{a}_i$ for fermions on lattice sites $i$~\cite{Nandkishore2015MBLReview,Imbrie2016}
\begin{equation}
\hat c_k = \sum_{i} u_i \hat a_i + \sum_{i,j,k} u_{ijk} \hat a_i \hat a_j^\dagger \hat a_k + \dots,
\end{equation}
such that the $l$-bit operators read ${\hat {n}}_k = \hat c_k^\dagger  \hat c_k $. In the MBL regime with strong disorder,
the coefficients $u_i$,  $u_{ijk}$ essentially have a local support with some exponential tails decaying on length scale $\xi$.
It can be shown~\cite{Giu18} that the terms that lead to particle-hole dressing $u_{ijk}$ and all higher order terms
contribute at least in first order in the interaction strength $V$, so for weak interactions $V$ and strong disorder the quasi-particles
are in leading order given by a single-particle transformation.


\subsection{MBL system in the presence of a phonon bath}

Let us now discuss the coupling of an MBL system  
to a phonon bath
with a  coupling operator $\hat{v}$
that is some single-particle operator $\hat{v}=\sum_{ij} v_{ij} \hat a_i^\dagger \hat a_j$,
\begin{align}
\hat{H}_\mathrm{SB}=\gamma \sum_{i,j} v_{ij} \hat a_i^\dagger \hat a_j \otimes \sum_{\alpha} \lambda_\alpha (\hat b_\alpha + \hat b_\alpha^\dagger).
\label{eq:sb-phonon-mbl}
\end{align}
The form of ${\hat H}_{SB}$ induces quantum jumps between eigenstates of $\hat{H}$ and the corresponding dissipator  
is given by
\begin{equation}\label{eq:heat-MBL-genrl}
{{\cal D}_{{\rm{heat}}}}\left( \hat{\rho}  \right) = \sum\limits_{\mathbf{n},\mathbf{n}'} \left( {\hat L}_{\mathbf{n}'{\bf n}}\hat{\rho} {\hat L}_{\mathbf{n}'{\bf n}}^\dag - \frac{1}{2} \left\{ {\hat L}_{\mathbf{n}'{\bf n}}^\dag {\hat L}_{\mathbf{n}'{\bf n}},\hat{\rho} \right\} \right),
\end{equation}
with $\hat L_{\mathbf{n}'{\bf n}} =\sqrt{R_{\mathbf{n}'{\bf n}}}|{\bf n}'\rangle \langle {{\bf n}}|$ describing the jump from Fock state
$|{\bf n}\rangle$ 
to the state $|{{\bf n}'}\rangle$. 
The associated jump rate is then  given by 
\begin{equation}\label{Rkq}
{{R_{{\bf n}'{\bf n}}}} = {2\pi }{\gamma ^2} \vert\bra{\mathbf{n}'} \hat{v} \ket{\mathbf{n}}\vert^2 g\left( {{E_{{{\bf n}'}}} - {E_{\bf n}}} \right),
\end{equation}
with  $g(E)$ the bath correlation function [Eq.~\eqref{eq:bathcorr}].
Note that generally, in the full spectrum of a quantum many body system
one expects that a large number of close degeneracies occur, which violates the validity condition in Eq.~\eqref{eq:cond-rotwave}.
However, for MBL systems  in the limit of weak interactions and strong disorder that we are aiming at,
we will find significant rates only for processes, where one or at most a few excitations are transferred.
Moreover, typically the bath will transfer excitations only between nearby $l$-bits, since $v_{ij}$ decays on the correlation lengths of the bath. Both together leads to a much milder condition.

For weak interactions and strong disorder, particle-hole dressing is suppressed, 
such that the transformation to the $l$-bits
effectively becomes a single-particle transformation
\begin{equation}\label{eq:trafo-noph}
	\hat c_k^\dagger \approx \sum_i \psi_k(i) \hat a_i^\dagger.
\end{equation}
In this case, the 
dissipator simplifies to
\begin{equation}\label{heat}
{{\cal D}_{{\rm{heat}}}}\left( \hat{\rho}  \right) = \sum\limits_{\mathbf{n},k,q} \left[ {{\hat L}_{qk}}\left( {\bf n} \right)\hat{\rho} {\hat L}_{qk}^\dag \left( {\bf n} \right) - \frac{1}{2} \left\{ {\hat L}_{qk}^\dag \left( {\bf n} \right){{\hat L}_{qk}}\left( {\bf n} \right),\hat{\rho} \right\} \right],
\end{equation}
with ${\hat L}_{qk}\left( {\bf n} \right) =\sqrt{R_{{\mathbf{n}_{qk}},{\bf n}}}|{{\bf n}_{qk}}\rangle \langle {{\bf n}}|$ describing the jump from Fock state
$|{\bf n}\rangle = \ket{n_1,\ldots,n_k,\ldots,n_q,\ldots,n_M}$ to the state $|{{\bf n}_{qk}}\rangle= \ket{n_1,\ldots,n_k-1,\ldots,n_q+1,\ldots,n_M}$ by transferring a particle from  the  single-particle mode $k$ to the mode $q$.
The fact that we neglect particle-hole dressing leads to
\begin{equation}
 \vert\bra{\mathbf{n}_{qk}} \hat{v} \ket{\mathbf{n}}\vert^2 = |{\nu_{kq}}|^2{n_k}^2\left( {1 - {n_q}} \right)^2 = |{\nu_{kq}}|^2{n_k}\left( {1 - {n_q}} \right),
\end{equation}
with the single-particle matrix element of the coupling operator
\begin{equation}
|\nu_{kq}|^2 =  \left |  \sum_{i,j} \psi^*_k(i) v_{ij} \psi_q(j) \right |^2.
\end{equation}

For convenience, we introduce an effective single-particle rate
\begin{equation} \label{eq:eff-sp-rate-Fock}
\tilde R_{qk} (\mathbf{n}) = {2\pi }{\gamma ^2} |\nu_{kq}|^2  g\left( {{E_{{{\bf n}_{qk}}}} - {E_{\bf n}}} \right),
\end{equation}
which in the case noninteracting case, with $E_{{\bf n}_{qk}} -E_{\bf n}=\varepsilon_q-\varepsilon_k$, 
 reduces to the single-particle rate $\tilde R_{qk} (\mathbf{n}) =R_{qk}$.
This allows for the extraction of quantum statistical factors from the many-particle rate
\begin{equation}\label{Rn}
{{R_{{{\bf n}_{qk}},{\bf n}}}} = \tilde R_{qk} (\mathbf{n}) \left( {1 - {n_q}} \right) {n_k}.
\end{equation}
This expression resembles the expression found for the ideal gas \cite{DanielPRE}, where
the many-particle rate is the single-particle rate $R_{qk}$ multiplied by the occupation of the departure state $n_k$
and the Pauli blocking factor $(1-n_q)$ of the target state. The difference is that due to interactions,
the  transition rate depends on the whole configuration, rather than
only on the two single-particle states involved in the transition.
%

\subsection{MBL system in the presence of a particle reservoir}

Let us now  turn to the case where the system may also exchange particles with an external fermionic reservoir
with temperature $T$ and chemical potential $\mu$. For simplicity, we again consider the regime where we can neglect particle-hole dressing.
Here the system-bath Hamiltonian is given by
\begin{align}
\hat{H}_\mathrm{SB}= \sum_{\alpha}  \lambda_\alpha (\hat b_\alpha \hat d^\dagger+ \hat b_\alpha^\dagger \hat d),
\label{eq:sb-fermion}
\end{align}
where $\hat{d} = \sum_i \varphi(i) \hat{a}_i$ is an arbitrary single-particle mode in the system. The resulting dissipator reads
\begin{equation}\label{gain}
{\cal D}_\mathrm{part} \left( \hat{\rho}  \right) =  \sum_{\alpha=\pm} \sum\limits_{\mathbf{n}, k}\left [ {{\hat L}_{\alpha,k}({\bf n})}\hat{\rho} {\hat L}_{\alpha,k}^\dag({\bf n})
- \frac{1}{2} \left\{{\hat L}_{\alpha,k}^\dag({\bf n}) {\hat L}_{\alpha,k}({\bf n}), \hat{\rho} \right\} \right ],
\end{equation}
where ${\hat L}_{+,k}({\bf n}) = \sqrt{\Gamma_{+,\bf n}}\ket{\mathbf{n}} \bra{\mathbf{n}_{k\downarrow}}$ and ${\hat L}_{-,k}({\bf n}) = \sqrt{\Gamma_{-,\bf n}}\ket{\mathbf{n}_{k\downarrow}} \bra{\mathbf{n}}$,
with $\ket{\mathbf{n}_{k\downarrow}} =\ket{n_1, \dots n_k-1, \dots}$
being the Fock state obtained by removing one particle from mode $k$. The jump rates $\Gamma_{{\pm},\bf n}$ are given by
\begin{equation}\label{Gamma}
\Gamma_{{\pm},\bf n} = \nu_k(\mathbf{n}) g_\mathrm{F}^{\pm}(E_\mathbf{n}-E_{\mathbf{n}_{k\downarrow}}),
\end{equation}
with the coupling matrix element
\begin{align}\label{nu_k}
	\nu_k(\mathbf{n}) = |\bra{\mathbf{n_{k\downarrow}}} \hat{d} \ket{\mathbf{n}}  |^2 = \left |\sum_i \varphi^*(i){\psi _k}(i) \right |^2n_k \equiv \eta(k) n_k,
\end{align}
and
the bath-correlation function for the particle exchange with the fermionic bath
\begin{align}
	g_\mathrm{F}^+ (E)&= J(E) f(E, \mu, T), \\
	g_\mathrm{F}^- (E)&=   J(-E) [1-f(-E, \mu, T)],
\end{align}
where
$ f(E,\mu, T) = (e^{(E - \mu)/T} + 1)^{-1}$ is the Fermi distribution of the bath.

\section{Interacting Anderson insulator as quantum wire}
\label{sec:model}

\begin{figure*}[h]
    \centering
    {\includegraphics[width=0.7\columnwidth]{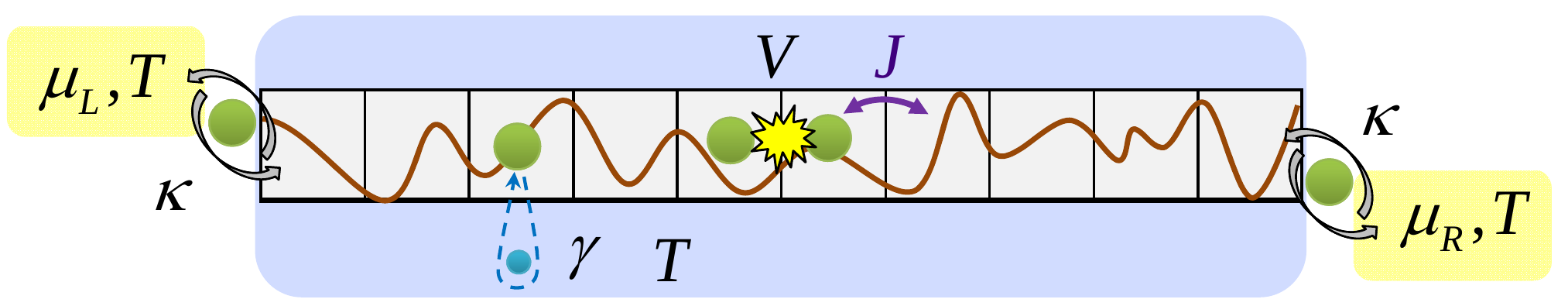}}
    \caption{Schematic illustration of the model under consideration. An open tight-binding chain of interacting spinless fermions (green dots)
    in a random potential (brown curve) is coupled to a global heat bath at temperature $T$ (blue box).
    This global bath induces phonon (blue dot)-assisted heat exchange with the system and two local baths (yellow boxes) at the ends with
    different chemical potentials ($\mu_L$ and $\mu_R$), which induce particle transport through the chain.
    }\label{model}
\end{figure*}

We benchmark the explained method studying a disordered extended Hubbard chain of spinless fermions (Fig.~\ref{model}). The system Hamiltonian~\cite{PhysRevB.75.155111,PhysRevB.77.064426,PhysRevB.82.174411} reads
\begin{equation}
\label{H}
 {\hat H} = {\hat H_0} + V \sum_{i=1}^{M-1} {\hat a}_i^\dag {{\hat a}_i}{\hat a}_{i+1}^\dag {\hat a}_{i+1},
\end{equation}
with the single-particle term given by
\begin{equation}\label{H0}
{{\hat H}_0} =  - J\sum\limits_{i=1}^{M-1} {\left( {{\hat a}_i^\dag {{\hat a}_{i + 1}} + {\hat a}_{i + 1}^\dag {{\hat a}_i}} \right)}  + \sum\limits_{i=1}^{M} {{w _i}{\hat a}_i^\dag {{\hat a}_i}},
\end{equation}
where ${\hat a}_i$ (${\hat a}_i^\dagger)$ is the fermionic annihilation (creation) operator at site $i$ in a chain of length $M$.
Moreover, $w_i$ are random fields uniformly distributed in the interval $[-W, W]$, $J$ and $V$ are respectively the hopping and the interaction strength.
${\hat H}_0$ describes the non-interacting Anderson-model~\cite{Anderson1958PhysRev.109.1492} (see appendix \ref{AppAnderson}) and all its single-particle eigenstates are exponentially localized for any amount of disorder $W$~\cite{Goldshtein1977}.
For $V\ne 0$ the model has an MBL transition~\cite{Altman2015MBLReview,Nandkishore2015MBLReview,ALET2018MBLReview,abanin2018ergodicity}.
At large disorder $W\gg V$, the system is in the MBL phase and it is a perfect insulator, while for weak disorder $W\ll V$ the system is ergodic, describing a thermal phase.
We couple the system to
two leads at the ends of the chain. The leads have the same temperature $T$ but different chemical potentials,
$\mu_{L}$ and $\mu_R$,  which exchange particles with the system,
inducing thus particle transport through the chain.
Moreover, the system is also coupled to a global thermal bath also at temperature $T$ (unless stated otherwise), with which it exchanges energy, as sketched in Fig.~\ref{model}.

We are interested in the limit of weak interactions at strong disorder, where in leading order $l$-bits  can be approximated
 by the non-interacting Anderson operators~\cite{Giu18}, so that we arrive at the approximate Hamiltonian
\begin{equation}\label{Happ}
{\hat H}_{\rm app} = \sum\limits_k {{\varepsilon _k}{{\hat n}_k}}  + \frac{1}{2}\sum\limits_{k,q} {{U _{kq}}{{\hat n}_k}{{\hat n}_q}}.
\end{equation}
Here ${\hat n}_k = {\hat c}^\dagger_k {\hat c}_k$ with ${\hat c}_k^\dagger = \sum_i \psi_k(i) {\hat a}_i^\dagger$, and  $\varepsilon_k$,  $\psi_k$  are the single-particle eigenenergies and eigenmodes, respectively.
The interaction
coefficients $U_{kq} $ are given by
\begin{equation}
U_{kq} =  -2V\sum_{i} [ \psi_k(i) \psi_q (i) \psi_k (i+1) \psi_q (i+1) - |\psi_k(i)|^2  |\psi_q(i+1)|^2 ].
\end{equation}
\begin{figure*}
    \centering
      \subfloat[][$W=2J$]{\includegraphics[width=0.33\columnwidth]{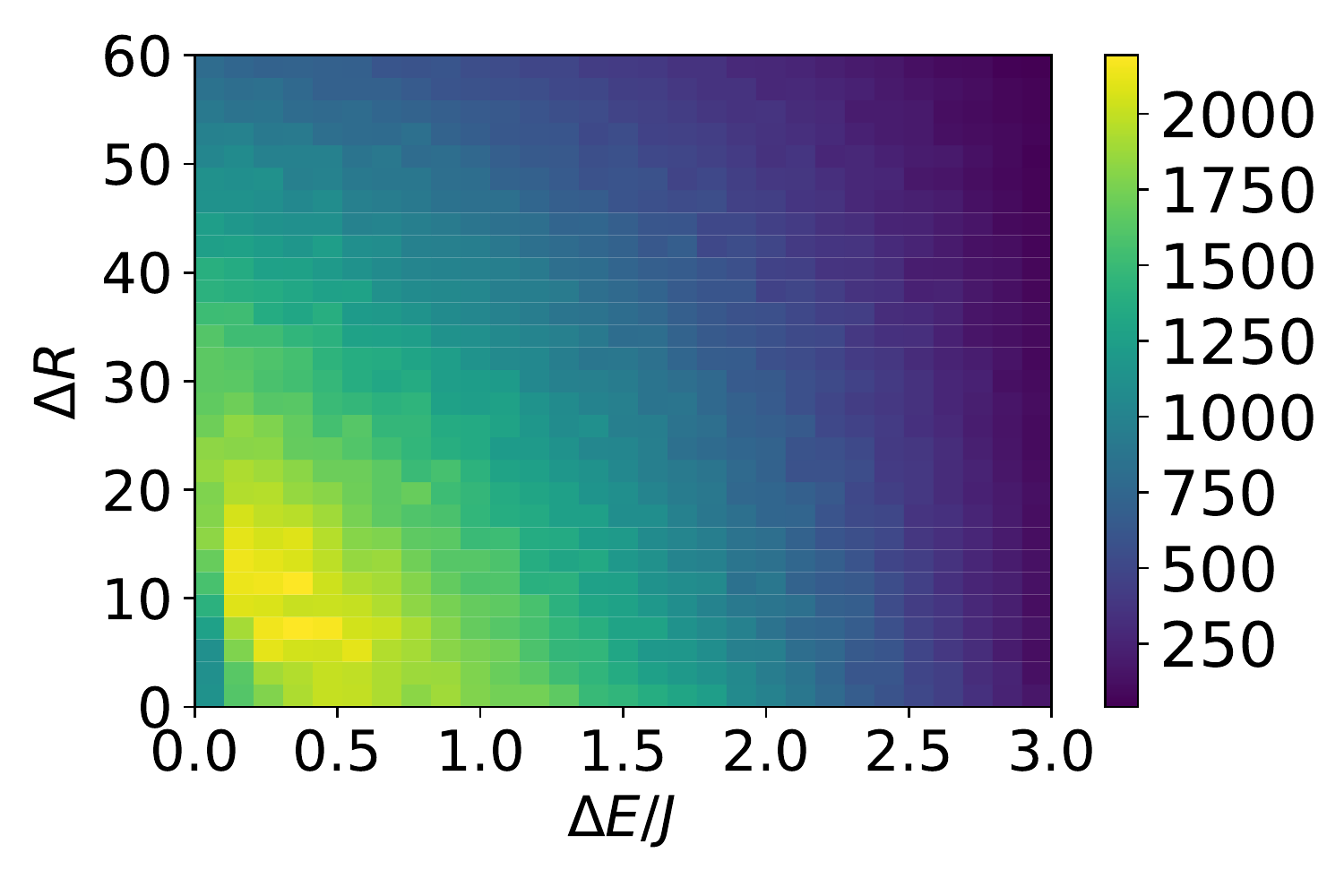}}
     \subfloat[][$W=3J$]{\includegraphics[width=0.33\columnwidth]{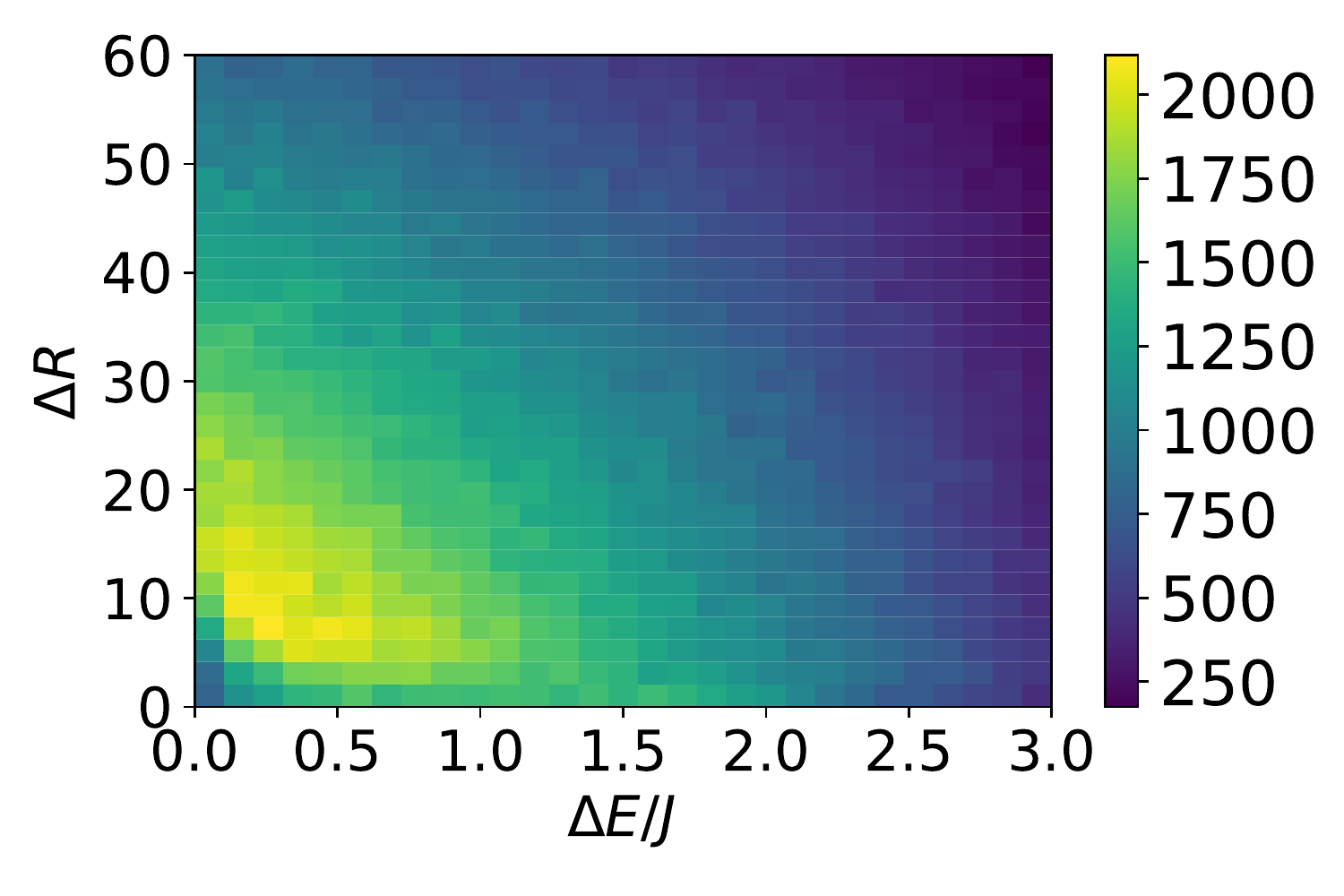}}
     \subfloat[][$W=5J$]{\includegraphics[width=0.33\columnwidth]{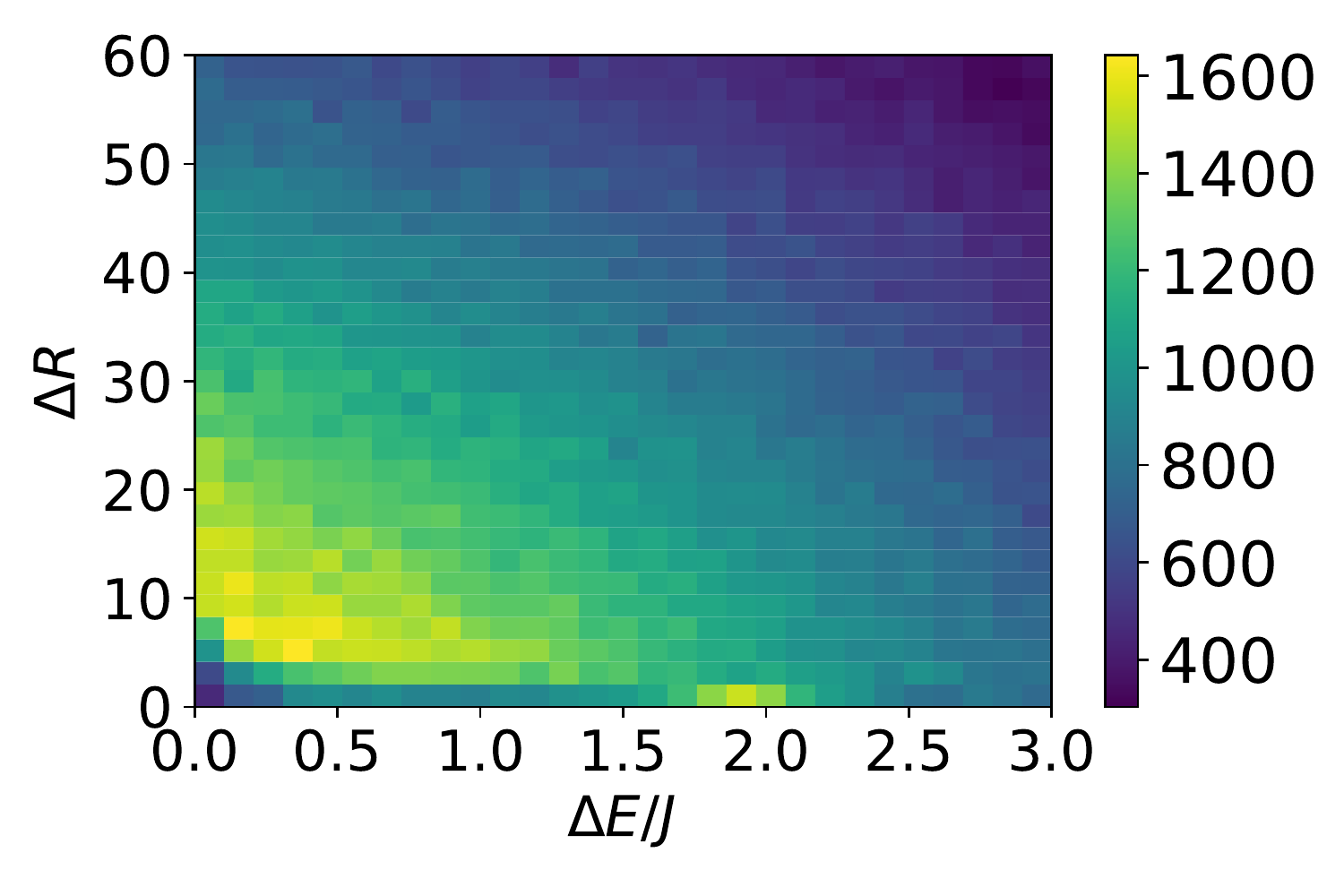}}\\
      \subfloat[][$W=2J$]{\includegraphics[width=0.33\columnwidth]{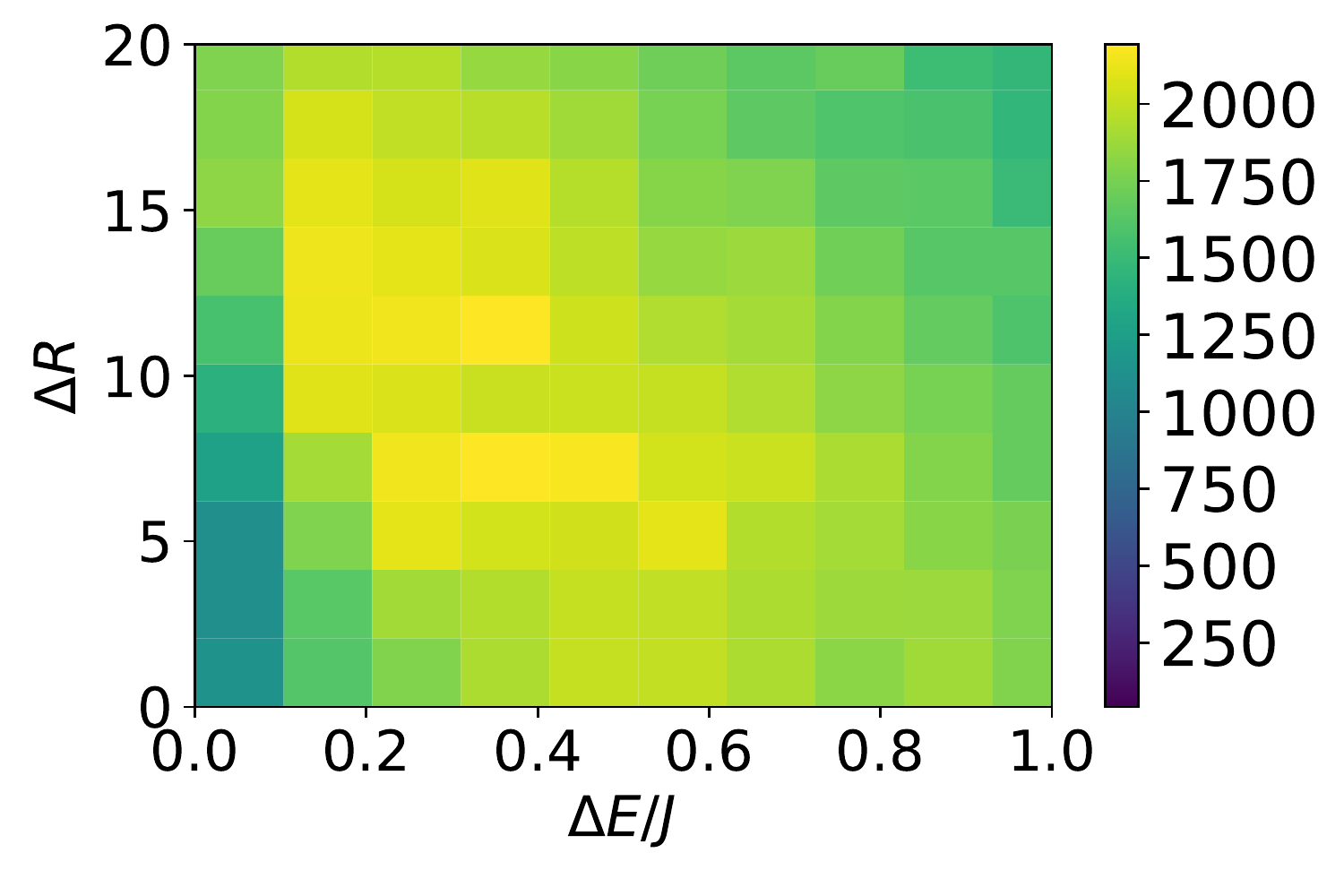}}
     \subfloat[][$W=3J$]{\includegraphics[width=0.33\columnwidth]{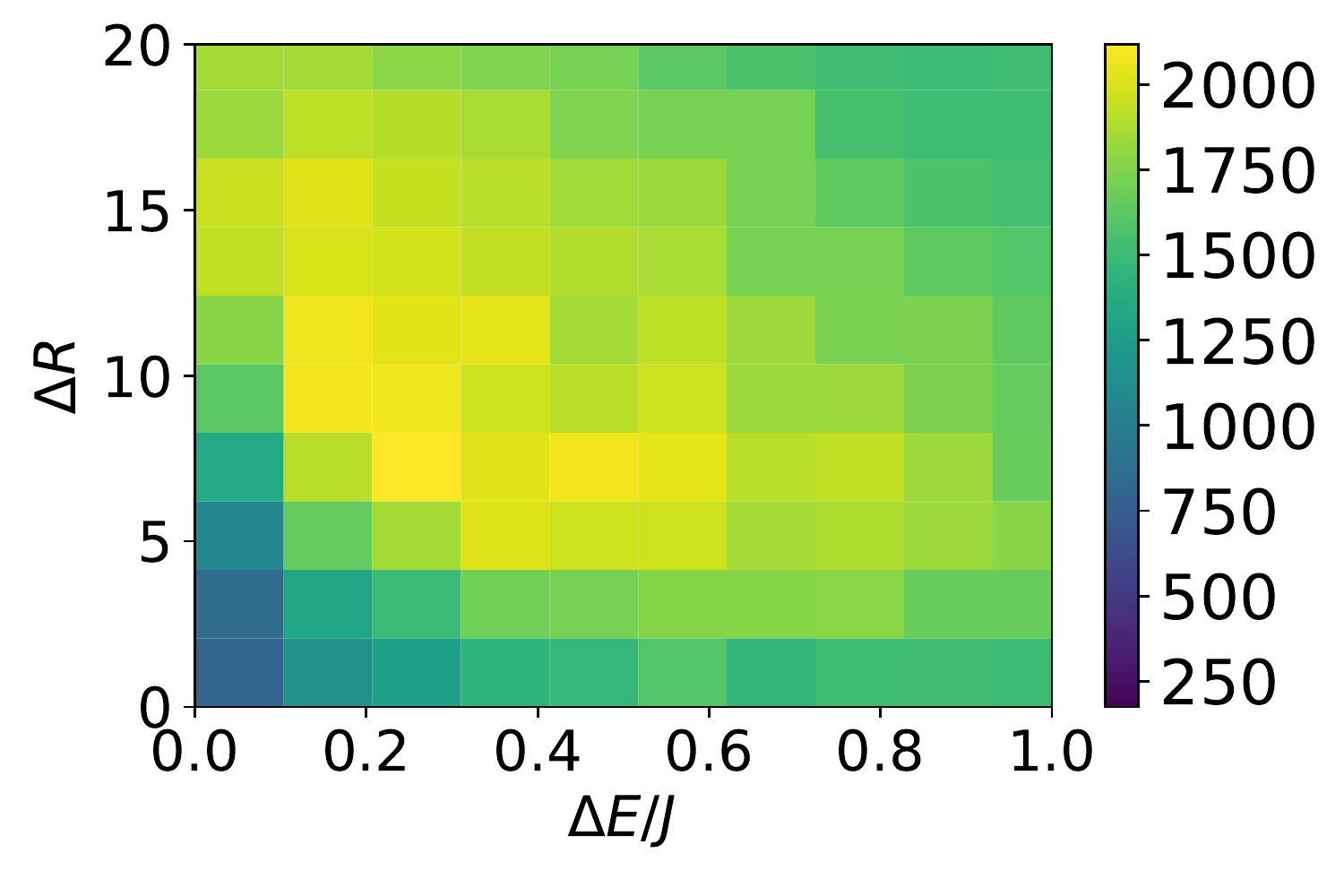}}
     \subfloat[][$W=5J$]{\includegraphics[width=0.33\columnwidth]{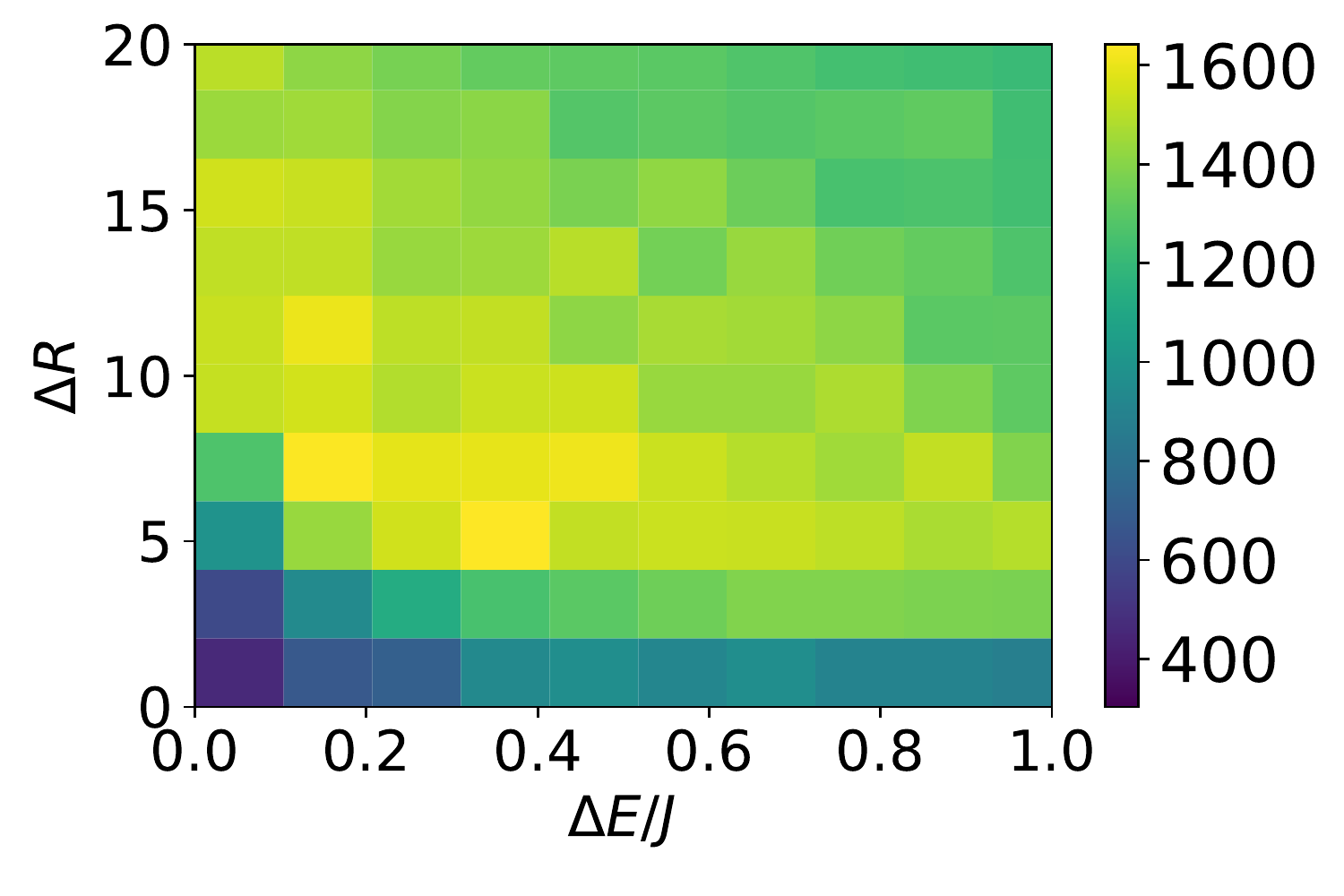}}
    \caption{
    Distribution of the spatial distance of two eigenstates $\Delta R$ as a function of their energy difference $\Delta E$. 1000 disorder realizations are added up.
    (d)-(f) are respectively the zoom-in views of (a)-(c).
    }\label{RE}
\end{figure*}
Since the orbitals are exponentially localized, $\psi_k(i) \propto e^{-|i-i_k|/\xi_{\text{loc}}(k)}$, with localization center $i_k$ and
the single-particle localization length $\xi_{\text{loc}}$~\cite{Kramer1993}, after a relabeling of the indexes $(k,q)$ according to their spatial positions,
we have $U_{kq}\sim V e^{-|k-q|/\xi_{\text{loc}}}$ (here $\xi_{\text{loc}}$ is the localization length in the middle of the single-particle spectrum).
This approximation [Eq.~(\ref{Happ})] is equivalent to discarding the off-diagonal elements of the full Hamiltonian [Eq.~(\ref{H})] in the non-interacting basis.
 Its reliability for weak interactions has been shown in Ref.~\cite{Giu18}.
Its validity is strongly supported also by the spatio-energetic anti-correlations between the single-particle Anderson orbitals.
If two single-particle eigenstates are close to each other in space, their energy difference is more likely to be large,
suppressing scattering events between these states.
Figure.~\ref{RE} gives evidence of the existence of this anti-correlation, showing the distribution of the spatial distance between pairs of eigenstates
$\Delta R = | \sum_i i (|\psi_k(i)|^2 -  |\psi_q(i)|^2) |$ as a function of their energetic distance $\Delta E = |\varepsilon_k-\varepsilon_q|$.
The distribution is mainly concentrated at the left bottom corner of Figures (a)-(c), which is zoomed in (d)-(f).
It is important to point out
that a better approximation for the $l$-bits can be obtained
systematically by treating the omitted matrix elements perturbatively,
giving rise to higher order  corrections in interaction strength~\cite{Fischer2016PhysRevLett.116.160401}.
Such a procedure is beyond the scope of this paper but it would not render our approach inefficient.


We couple this system to a heat reservoir via the system-bath Hamiltonian
\begin{equation}\label{HSB}
 {{\hat H}_{SB}} = \gamma \sum\limits_i{ {\hat n}_i \otimes  \sum\limits_\alpha  {{C_\alpha }\left( {{\hat b}_\alpha ^{(i)\dag}  + {{\hat b}_\alpha^{(i)} }} \right)}},
\end{equation}
which couples each site to an independent bath of bath-correlation length zero.
$C_\alpha$ and $\gamma$ are the coupling strengths.
Furthermore, we assume an Ohmic bath, with spectral density $J(E)$ proportional to $E$, $J(E) \propto E$.
The total rate that enters in Eq.~\eqref{heat} is therefore the sum of the rates for individual heat baths that couple to a single site $i$ only.

We investigate the transport properties of the steady state, when the system is coupled to the two leads and the thermal bath.
The lead coupling is described by
\begin{align}
 \hat H_\mathrm{SB, L} &=  \sqrt{\kappa} \sum_{\alpha}  D_\alpha (\hat b^{(L)}_\alpha \hat a^\dagger_1+ \hat b^{(L)\dagger}_\alpha \hat a_1),\\
 \hat H_\mathrm{SB, R}&= \sqrt{\kappa} \sum_{\alpha}  D_\alpha (\hat b^{(R)}_\alpha \hat a^\dagger_M+ \hat b^{(R)\dagger}_\alpha \hat a_M).
\end{align}
The resulting coupling rates $\eta_L(k) = |{\psi _k}(1){|^2}$, $\eta_R(k) = |{\psi _k}(M){|^2}$
that enter in Eq.~\eqref{nu_k} are proportional to the value of the wavefunctions 
at the ends of the chain. 
For the leads we assume a constant density of states.

The effective single-particle energies 
are shifted due to interactions.
Within the scope of the approximate Hamiltonian in Eq.~\eqref{Happ}, it is convenient to define an interaction-shifted energy operator for each $l$-bit
\begin{equation}\label{Ek}
{\hat {\tilde{\varepsilon}}}_k = \varepsilon_k + \sum\limits_q {U_{kq} {\hat n}_q}.
\end{equation}
The energy difference occurring in the rate for the heat bath, Eq.~\eqref{eq:eff-sp-rate-Fock}, can then be expressed as
\begin{equation}\label{dE}
{E_{{{\bf n}_{qk}}}} - {E_{\bf n}} = \langle {{\bf n}} |{\hat { \tilde{\varepsilon}} _q} | {{\bf n}} \rangle  - \langle {{\bf n}} | {\hat { \tilde{\varepsilon}} _k} |{{\bf n}} \rangle  - {U_{kq}}.
\end{equation}
For the particle reservoir, the energy difference in Eq.~\eqref{Gamma} is given by
\begin{align}
	E_\mathbf{n}  - E_{\mathbf{n}_{k\downarrow}}= \langle {{\bf n}} | {\hat { \tilde{\varepsilon}} _k} |{{\bf n}} \rangle .
\end{align}
Note that writing energy differences  in terms of $\hat{\tilde\varepsilon}_k$ does not constitute an additional approximation.

\section{Solving the Master equation}
We employ two methods to compute the non-equilibrium steady state of  the master equation.
 The first method is a quantum-jump Monte Carlo
technique, which makes use of the fact that the equations of motion for the Fock-space occupation probabilities $p_\mathbf{n}$
can be mapped to a classical random walk. This method gives accurate results after sufficient statistical sampling.
The second method consists in deriving kinetic equations of motion by employing a mean-field decomposition of density-density correlations.
Importantly, both approaches are found to agree very well.




\subsection{Method I: Quantum-jump Monte Carlo simulation}
\label{sec:montecarlo}
As we discuss in Section~\ref{sec:Born-Markov}, the dynamics of the many-body occupation probabilities $p_{\bf n} = \langle {\bf n} | \hat{\rho} | {\bf n} \rangle$ decouples from the off-diagonal elements,
which decay over time. For our model, the equation of motion for the probabilities is given by
\begin{align}\label{pn}
\begin{split}
{\partial _t}{p_{\bf n}} = \sum\limits_{k,q}  \left( {1 - {n_q}} \right) {n_k} &\left[ \tilde R_{kq} (\mathbf{n}_{qk}){p_{{{\bf n}_{qk}}}} - \tilde R_{qk} (\mathbf{n}){p_{\bf n}} \right]  \\
+ \sum\limits_{\alpha=L,R}\sum\limits_k \kappa \eta_\alpha(k) & \left\{f_\alpha(\tilde{\varepsilon}_k) \left[ {{n_k}{p_{{{\bf n}_{k \downarrow }}}} - \left( {1 - {n_k}} \right){p_{\bf n}}} \right] \right. \\
  &\left. +{(1-f_\alpha(\tilde{\varepsilon}_k))\left[ {\left( {1 - {n_k}} \right){p_{{{\bf n}_{k \uparrow }}}} - {n_k}{p_{\bf n}}} \right]}\right\} ,
\end{split}
\end{align}
where ${\tilde{\varepsilon}}_k = \langle {\bf n}|{\hat {\tilde{\varepsilon}}}_k |{\bf n}\rangle$,
and we use the short notation $f_\alpha(E)=f(E, \mu_\alpha, T_\alpha)$ with $\alpha=L,R$.

The first line in Eq.~(\ref{pn}) describes the heat exchange processes, with the first term denoting the increase of the probability $p_{\bf n}$
by jumping from  $|{\bf n}_{qk}\rangle$ to the state $|{\bf n}\rangle$, and the second term denoting the inverse process.
The second line in Eq.~(\ref{pn}) describes particle exchange with the leads, with the first term denoting particle gain, and the second term the particle loss.

We generalize the quantum-jump Monte-Carlo method described in Ref.~\cite{Andrew2014AP} to the case of number-dependent single-particle rates $\tilde R_{qk} (\mathbf{n})$.
The dynamics of the occupation probability
$p_{\bf n}$ is simulated by taking random walks in the classical space composed by the Fock states $|{\bf n}\rangle = |n_1,\ldots,n_k,\ldots,n_M \rangle$
(not their superpositions)~\cite{DanielPRE} corresponding to the given jump rates.
In our case, we have two types of jumps, the first one due to the global thermal bath and the second one due to the leads.
For the heat exchange processes, a jump transfers one particle from one single-particle eigenstate to another,
which conserves the total particle number. For the particle exchange process, a jump adds (or removes) a particle to (or from) one eigenstate.
We perform these simulations by using a Gillespie-type algorithm (see appendix \ref{QJMC}).
We can then compute steady state expectation values (e.g. $\langle \hat{n}_k\rangle, \langle\hat{n}_k\hat{n}_q\rangle$), by averaging over the long-time dynamics of many trajectories.

\subsection{Method II: Kinetic theory}
\label{sec:kinetic}

In order to treat larger systems and also to gain some intuitive understanding of the non-equilibrium steady state (NESS), we  will now derive the kinetic equations of motion for the mean occupation numbers $\langle {\hat n}_l\rangle$.
The time evolution of $\langle {\hat n}_l\rangle$, again for the special case of our model system, is given by
\begin{equation}\label{nl}
\frac{d}{{dt}}\langle {{\hat n}_l}\rangle  =  {\rm{tr}}\left[ {{{\hat n}_l}{\partial _t}\hat{\rho} } \right] = \left( \frac{d}{{dt}}\langle {{\hat n}_l}\rangle\right)_{{\rm{heat}}} +\left( \frac{d}{{dt}} \langle {{\hat n}_l}\rangle\right)_{{\rm{part}}} ,
\end{equation}
with (see appendix \ref{derivation})
\begin{equation}\label{dnhe}
\left(\frac{d}{{dt}}\langle {{\hat n}_l}\rangle\right) _{{\rm{heat}}}=\sum\limits_k\left[ \langle \hat{\tilde R}_{lk}{\hat n}_k(1-{\hat n}_l)\rangle -  \langle \hat{\tilde R}_{kl}{\hat n}_l(1-{\hat n}_k)\rangle \right],
\end{equation}
and
\begin{align}\label{dnpe}
\begin{split}
\left(\frac{d}{{dt}}\langle {{\hat n}_l}\rangle\right)_{\rm{part}} &=  \sum\limits_{\alpha=L,R} {\eta_\alpha(l)\left[ \langle f_\alpha(\hat{\tilde{\varepsilon}}_l) (1- {\hat n}_l)\rangle -\langle(1-  f_\alpha(\hat{\tilde{\varepsilon}}_l) ){\hat n}_l \rangle \right] }\\
&=  \sum\limits_{\alpha=L,R} {\eta_\alpha(l)  \langle f_\alpha(\hat{\tilde{\varepsilon}}_l) - {\hat n}_l \rangle }.
\end{split}
\end{align}
In Eq.~\eqref{dnhe}, we have defined an effective rate operator
\begin{equation}\label{oRqk}
	\hat{\tilde R}_{lk}  = \sum\limits_{\bf n}{	\tilde R_{lk} (\mathbf{n}) |{\bf n}\rangle \langle {\bf n}|}.
\end{equation}
Using Eq.~\eqref{dE}, we find
\begin{align}
	\hat{\tilde R}_{lk}  = {2\pi }{\gamma ^2} |\nu_{kl}|^2  g\left({\hat { \tilde{\varepsilon}} _l} -  {\hat { \tilde{\varepsilon}} _k} - {U_{lk}}\right).
\end{align}
The effect of interactions is to modify the energy difference between the states
($\hat{\tilde{\varepsilon}}_l - \hat{\tilde{\varepsilon}}_k -U_{l k}$), implying that the  transition rate depends on the whole configuration, rather than
only on the two single-particles states involved in the transition ($\varepsilon_l - \varepsilon_k$).

In order to obtain a closed set of equations in terms of mean occupation values, we employ the mean-field approximation
\begin{align}
 \langle  \prod_{k=1}^s {\hat{n}_{i_k}} \rangle &\approx \prod_{k=1}^s \langle  {\hat{n}_{i_k}} \rangle, \quad \forall s, \text{ for } i_k \neq i_q, \forall k\neq q,
\end{align}
giving  $\langle f_\alpha(\hat{\tilde{\varepsilon}}_l) \rangle \approx f_\alpha(\langle \hat{\tilde{\varepsilon}}_l \rangle)  $,
$\langle g( \hat{\tilde{\varepsilon}}_l - \hat{\tilde{\varepsilon}}_k - \gamma_{l k} ){\hat n}_k(1-{\hat n}_l)\rangle
\simeq  g(\langle  \hat{\tilde{\varepsilon}}_l - \hat{\tilde{\varepsilon}}_k - \gamma_{l k} \rangle)\langle {\hat n}_k \rangle \langle 1-{\hat n}_l\rangle$.
In this way, we get a set of nonlinear kinetic equations of motion
{
\begin{equation}\label{dnndt}
\frac{d}{{dt}}\langle {\hat{n}_l}\rangle = \sum\limits_k {\left[ \langle{{\hat{\tilde R}_{lk}}\rangle \langle {\hat{n}_k}\rangle \left( {1 - \langle {\hat{n}_l}\rangle } \right)- \langle{\hat{\tilde R}_{kl}} \rangle\langle {\hat{n}_l}\rangle \left( {1 - \langle {\hat{n}_k}\rangle } \right) } \right]}  + \sum\limits_{\alpha=L,R} { \eta_\alpha(l) \left[f_\alpha(\langle \hat{\tilde{\varepsilon}}_l \rangle)  - \langle \hat{n}_l \rangle\right]},
\end{equation}
We obtain the equations for the NESS, by setting Eq.~(\ref{dnndt}) to zero,
\begin{equation}\label{ss}
\frac{d}{{dt}}\langle {\hat n}_l\rangle_{\rm {NESS}}=0.
\end{equation}
In the following, we will always consider the steady state and, therefore, drop the subscript {NESS}.
Using the single-particle wavefunctions $\psi_l$ we can find the mean occupation number (density profile) in real space
\begin{equation}\label{ni}
\langle \hat{n}_i \rangle = \sum\limits_{l=1}^{M}{|\psi_l(i)|^2\langle \hat{n}_l \rangle}.
\end{equation}

The nonlinear term in Eq.~(\ref{dnndt}) prevents us from finding an analytic solution. However, in the
absence of the global thermal bath ($\gamma = 0$), the solution of Eq.~(\ref{ss}) is given by
\begin{equation}\label{n0}
\langle \hat{n}_l \rangle_{\gamma=0} = \frac{\eta_L(l)f_L(\langle\hat{\tilde{\varepsilon}}_l\rangle)+\eta_R(l)f_R(\langle\hat{\tilde{\varepsilon}}_l\rangle)}{\eta_L(l)  + \eta_R(l)}.
\end{equation}

\begin{figure*}
    \centering
     \subfloat[][$V=0,  W = J$]{\includegraphics[width=0.45\columnwidth]{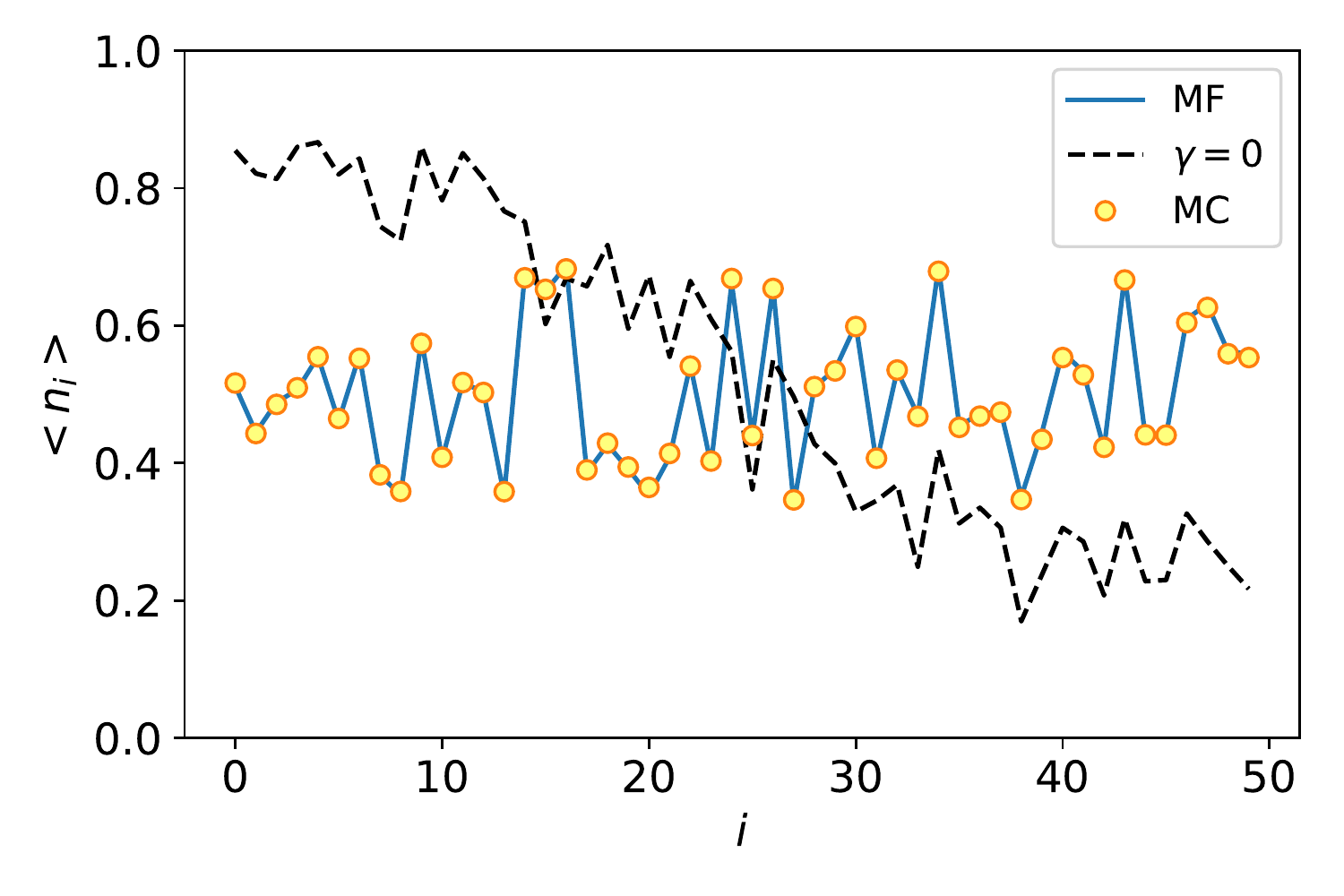}}
     \subfloat[][ $V=0,  W = 3J$]{\includegraphics[width=0.45\columnwidth]{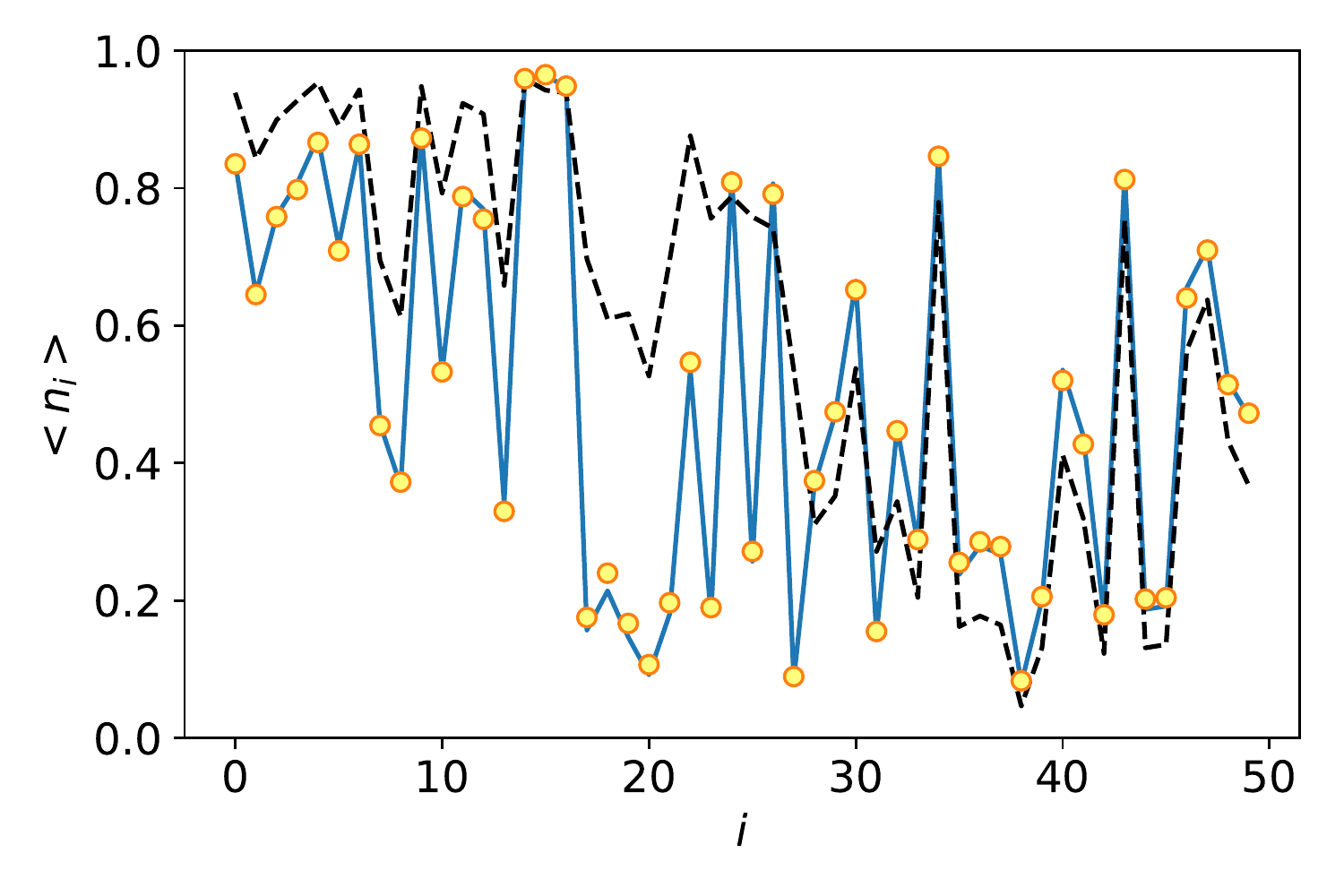}}\\
      \subfloat[][ $V=J,  W = J$]{\includegraphics[width=0.45\columnwidth]{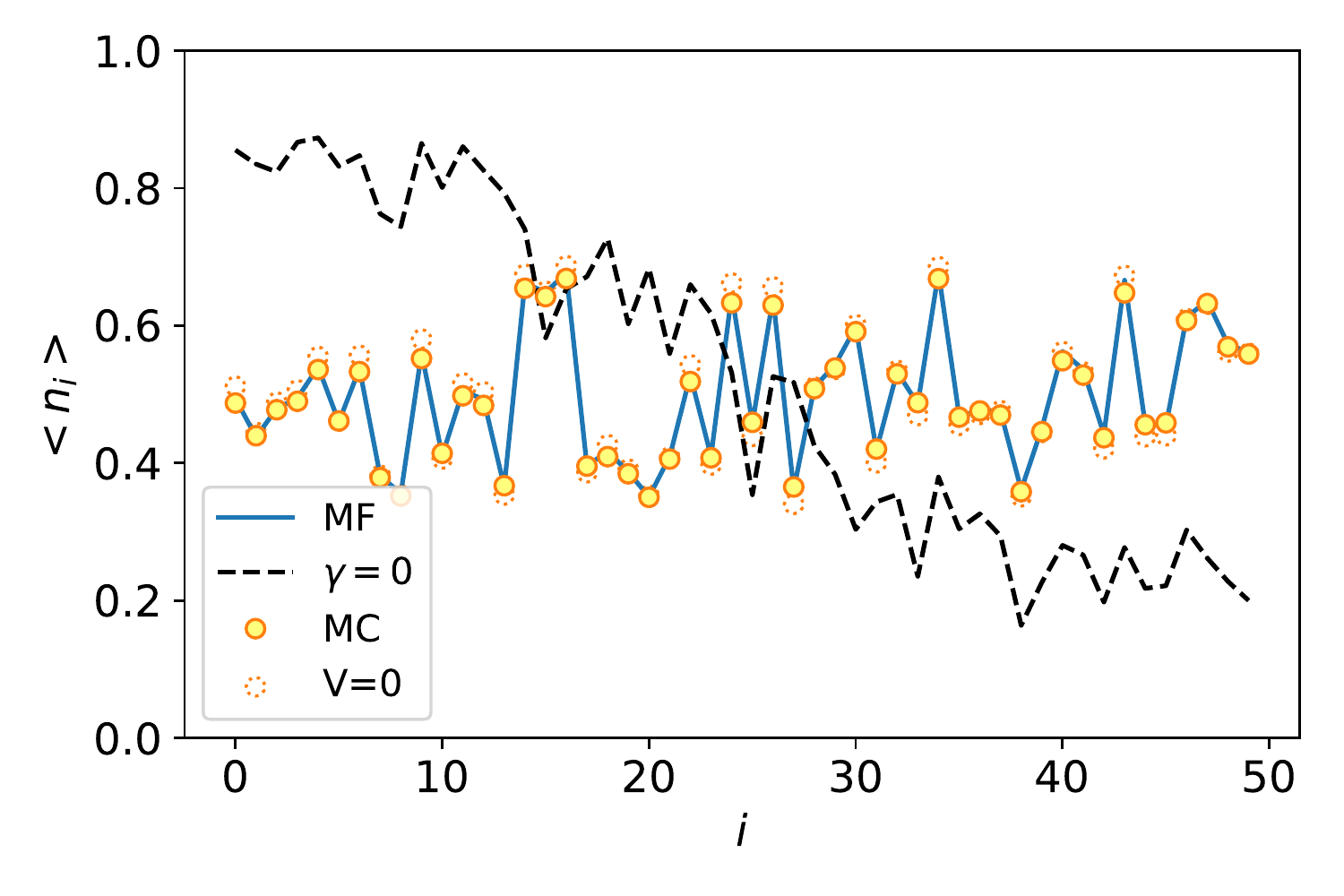}}
        \subfloat[][$V=J,  W =3 J$]{\includegraphics[width=0.45\columnwidth]{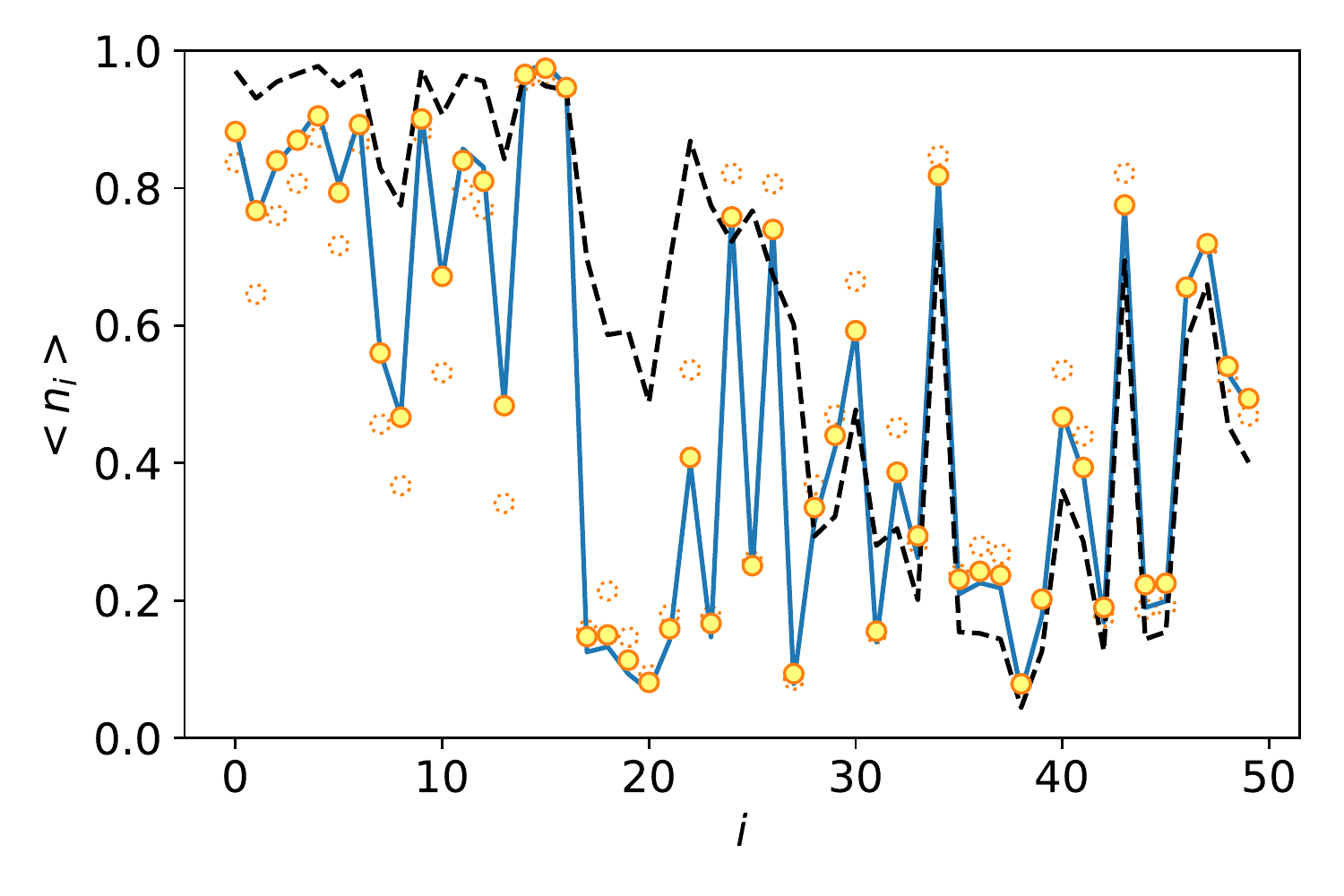}}
    \caption{ Density profile $\langle \hat{n}_i \rangle$ as a function of site $i$. The solid lines are the  results from the kinetic theory and the dashed lines are  the results
    in the absence of the global thermal bath (obtained by substituting Eq.~(\ref{n0}) into \eqref{ni}). The markers denote the averaged results over $1000$ trajectories of quantum-jump Monte-Carlo  (MC) simulations.
    The open circles in (c), (d) are the results with $V=0$, as shown in (a), (b).
    The parameters are $M = 50$,  $\gamma^2=0.01J$, $\kappa  = 0.1\gamma^2 $, $T=J$ and $\mu_L = \bar{\mu} +\delta\mu/2$, $\mu_R = \bar{\mu} -\delta\mu/2$,  with $\delta\mu=5J$, and $\bar \mu$ is set to make $N = M/2 = 25$.
    }\label{Ni}
\end{figure*}

Let us compare the  methods described in Sec.~\ref{sec:montecarlo} and  \ref{sec:kinetic} by computing the density profile $\langle \hat{n}_i \rangle$, which also determines the current to be discussed later.
Figure~\ref{Ni} shows the density profile $\langle \hat{n}_i \rangle$ obtained by numerically solving  the kinetic equations [Eq.~(\ref{ss}), solid lines]
and the one calculated using the quantum-jump Monte-Carlo technique [Eq.~(\ref{pn}), markers].
The two results are almost indistinguishable both for the non-interacting and the interacting case.
The chemical potentials in the leads are taken equal to $\mu_{L,R} = \bar \mu \pm \delta\mu/2$ with relatively large chemical potential difference $\delta\mu = 5J$, and
$\bar \mu$ is fixed so that the system is at half-filling ($\langle \hat{N}\rangle= M/2$, where $\hat{N}$ counts the total number of particles).
Later on, we will use only small chemical potential offsets $\delta \mu$, in order to compute conductivities.
The large difference between the chemical potentials produces a density gradient in the absence of the global thermal bath ($\gamma = 0$), as shown in Fig.~\ref{Ni} (black dashed lines).
The presence of the global thermal bath, which induces phonon-assisted tunneling, erases this gradient for weak disorder, as shown in (a) and (c).
Furthermore,  by increasing the disorder strength $W$, which prompts the localization of the wavefunctions and thus the suppression of particle tunneling, the density gradient is restored,
as shown in Fig.~\ref{Ni}(b) and (d).


\section{Results I: Observation of variable-range hopping for noninteracting Fermions in a microscopic model}
We investigate the particle current in the NESS, which is defined as the net particle flow at the ends of the chain
\begin{equation}\label{I}
I =\sum\limits_{l=1}^{M} { \eta_L(l) \left[f_L(\langle \hat{\tilde{\varepsilon}}_l \rangle)  - \langle \hat{n}_l \rangle\right]} = - \sum\limits_{l=1}^{M}{ \eta_R(l) \left[ f_R(\langle \hat{\tilde{\varepsilon}}_l \rangle)  - \langle \hat{n}_l \rangle\right]}.
\end{equation}
Having shown that the kinetic equations, Eq.~\eqref{dnndt}, are reliable,
we now use this approach to study
the dependence of the current $I$, Eq.~(\ref{I}), on disorder strength $W$,
on the temperature of the bath $T$, and on interaction strength $V$.
As shown in Appendix \ref{comparison}, for the current, the results of the kinetic equations also agree very well with those
from Monte-Carlo simulation. In the following, we will only show the results of the kinetic theory with disorder averaged.

\subsection{Transport as a function of  disorder strength $W$} 
 \begin{figure*}
    \centering
\subfloat[][]{\includegraphics[width=0.5\columnwidth]{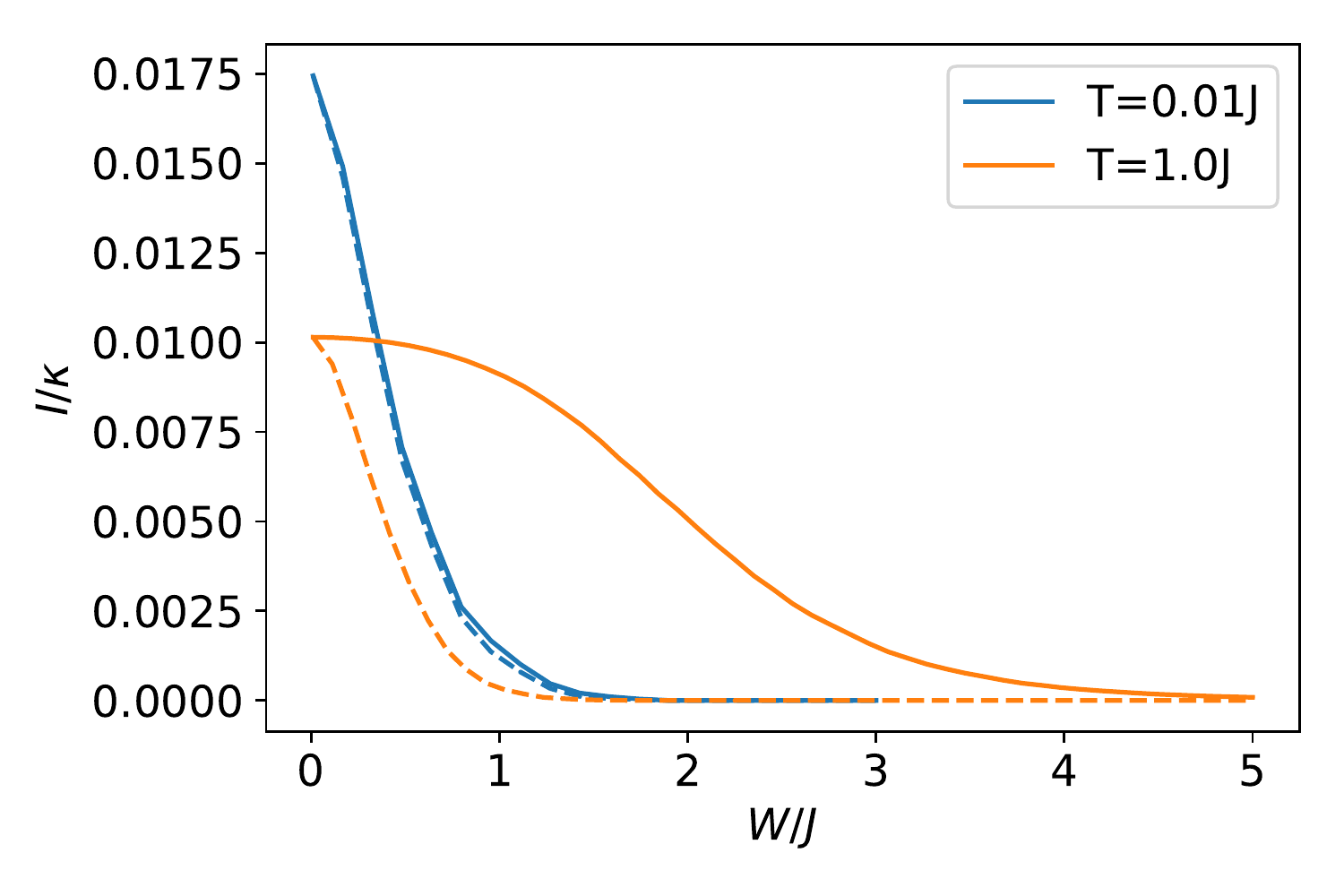}}
\subfloat[][]{\includegraphics[width=0.5\columnwidth]{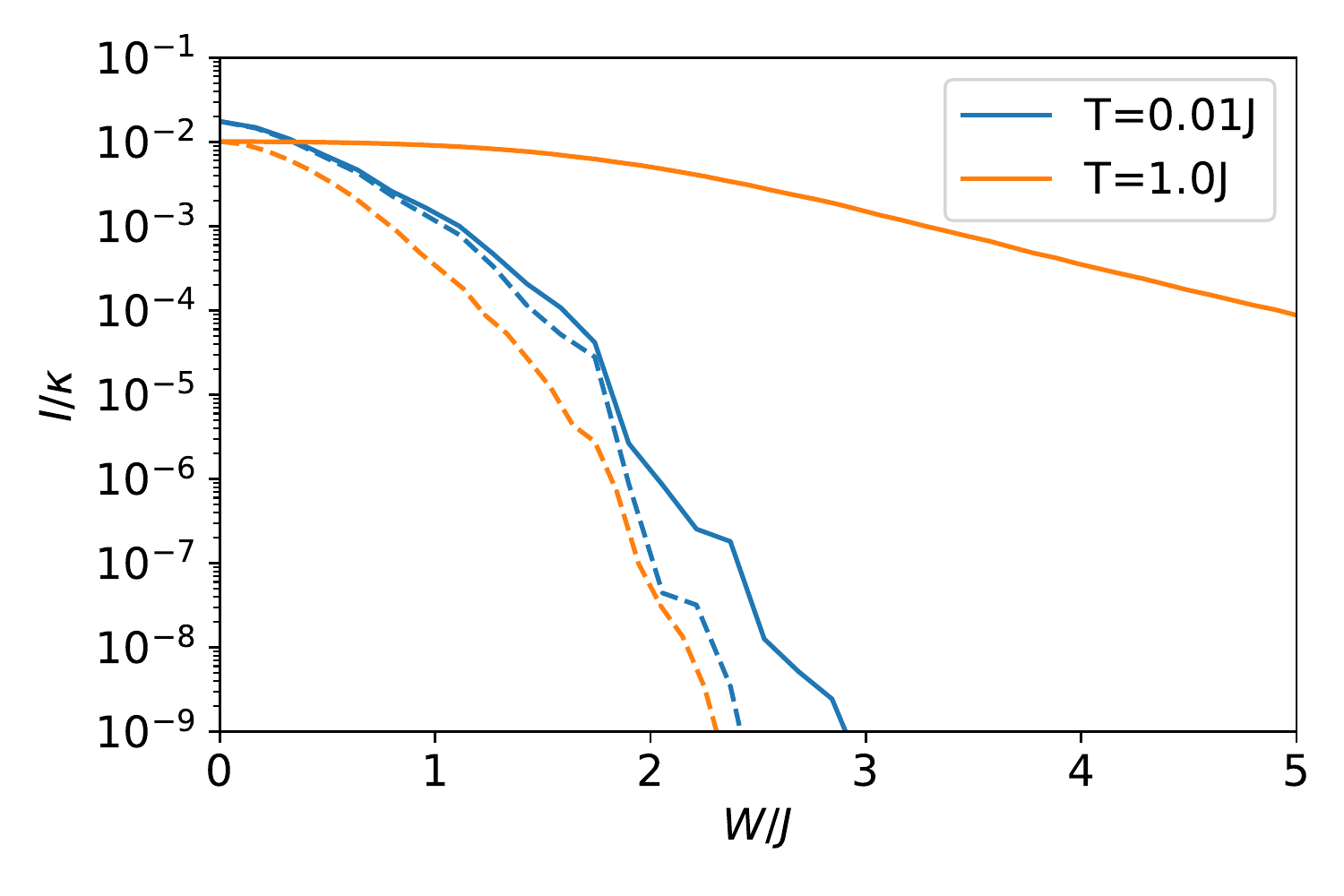}}
    \caption{Current $I$ as a function of the disorder strength $W$. The results are averaged over $1000$ disorder realizations.
    The solid lines are obtained from the kinetic equations and the dashed lines are the  results without coupling the system to the global bath [Eq.~(\ref{I0})]. The parameters are system size $M=100$,
    dissipation rate $\kappa = 0.1\gamma^2$, interaction strength $V = 0$, and $\mu_L = \bar{\mu} +\delta\mu/2$, $\mu_R = \bar{\mu} -\delta\mu/2$,
    with chemical potential imbalance $\delta \mu = 0.1J$, and $\bar \mu$ is set to make $N = M/2 = 50$. Except when otherwise stated, we will take the same parameters for the rest of the work.
    }\label{Wapp}
\end{figure*}
In the absence of the thermal bath ($\gamma = 0$), the current in the NESS is given by
\begin{equation}\label{I0}
I_{\gamma = 0} = \sum\limits_{l=1}^{M}{\left[ f_L(\langle \hat{\tilde{\varepsilon}}_l \rangle) -f_R(\langle \hat{\tilde{\varepsilon}}_l \rangle) )\right]\frac{\eta_L(l)\eta_R(l)}{\eta_L(l)+ \eta_R(l)}},
\end{equation}
which is obtained by substituting the solution of the occupation numbers Eq.~(\ref{n0}) into Eq.~(\ref{I}).
The terms of $I_{\gamma = 0}$ are proportional to the
product $\eta_L(l)\eta_R(l) \propto |\psi_l(1)|^2 |\psi_l(M)|^2 \sim \exp({-4 M/\xi_\text{loc}})$, where for strong disorder $\xi_{\text{loc}} \sim 1/\log{(W)}$. This implies that without coupling to a thermal environment the current will dramatically decay with { disorder strength}.
This is confirmed in Fig.~\ref{Wapp} (dashed lines), where the behavior of the current is shown as a function of $W$ for the
non-interacting chain of $M=100$ lattice sites. Here and in the following, we consider the weak chemical potential difference $\delta\mu=0.1J$.
Figure~\ref{Wapp} also reports  the case in which the thermal bath is present $(\gamma \ne 0)$, showing
that the presence of a global thermal bath enhances transport for sufficiently large $T$.

\subsection{Transport as a function of temperature $T$}
In this Section we study the dependence of the current $I$ on temperature $T$ and disorder strength $W$ for the noninteracting case.
Figure~\ref{T_fit}(a) shows the current $I$ as a function of $T$ for a fixed system size ($M=100$) for several values of $W$,
while Fig.~\ref{T_fit}(b) shows $I$ for a fixed disorder strength $W$ for several system sizes $M$.
We can distinguish three distinct regimes according to the value of the temperature.
At `low' temperatures~$T$, 
 the current approaches the result obtained in the absence of the global bath ($\gamma = 0$, dotted lines).
In the intermediate temperature regime, the behavior of the current is well predicted by Mott's law for variable-range hopping (VRH)~\cite{Mott1969}, $I \propto \exp({-\sqrt{T_0/T}})$.
Finally, at `high' temperatures ($T\gg J$), $I$ decreases as a function of $T$.

Note that the locations of  the three regions depend on various parameters,
such as the disorder strength $W$ (as shown in Fig~\ref{T_fit}(a)), the coupling to the leads $\kappa$, and so on.
\begin{figure*}
    \centering
     \subfloat[][$M=100$]{\includegraphics[width=0.5\columnwidth]{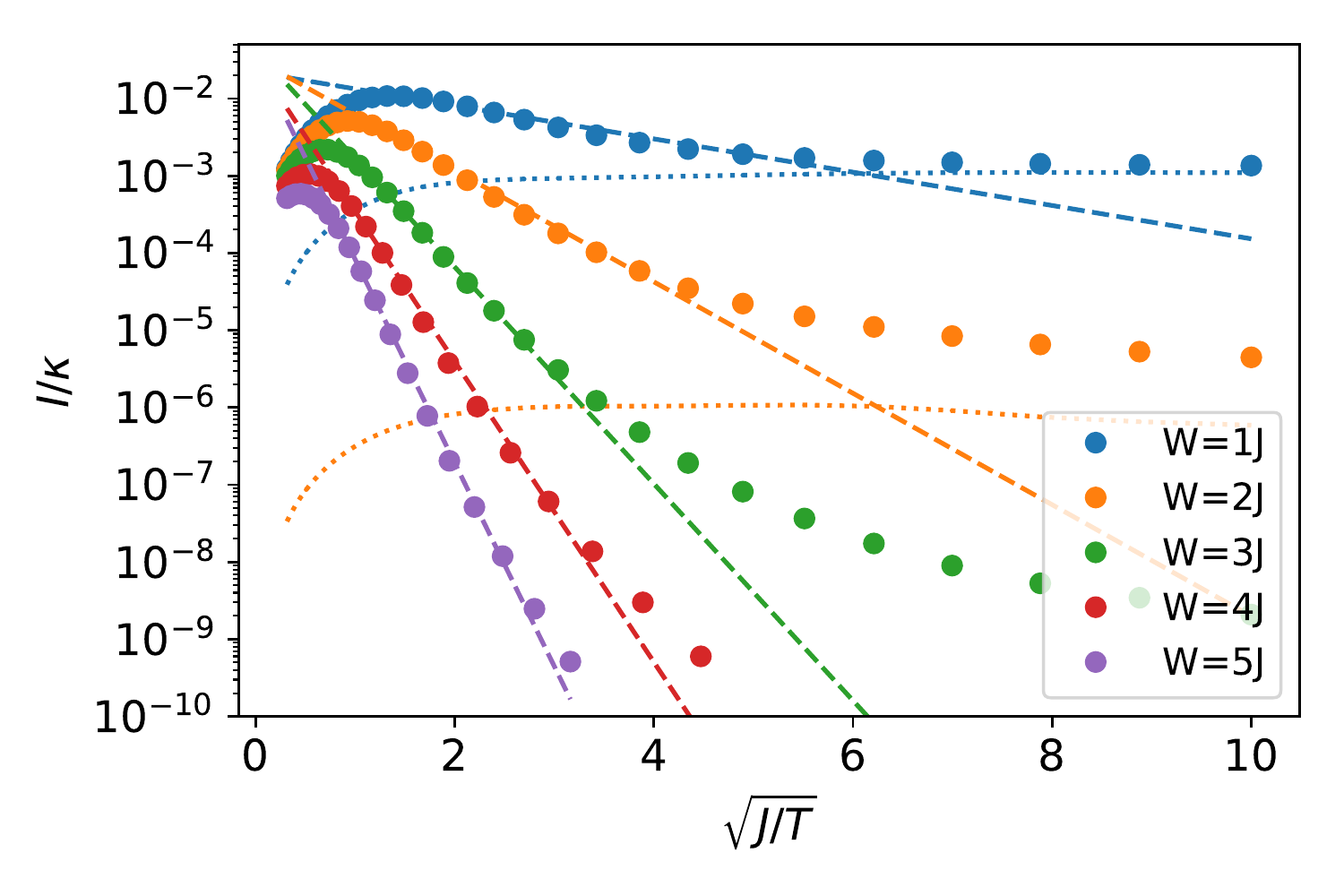}}
      \subfloat[][$W=3J$]{\includegraphics[width=0.5\columnwidth]{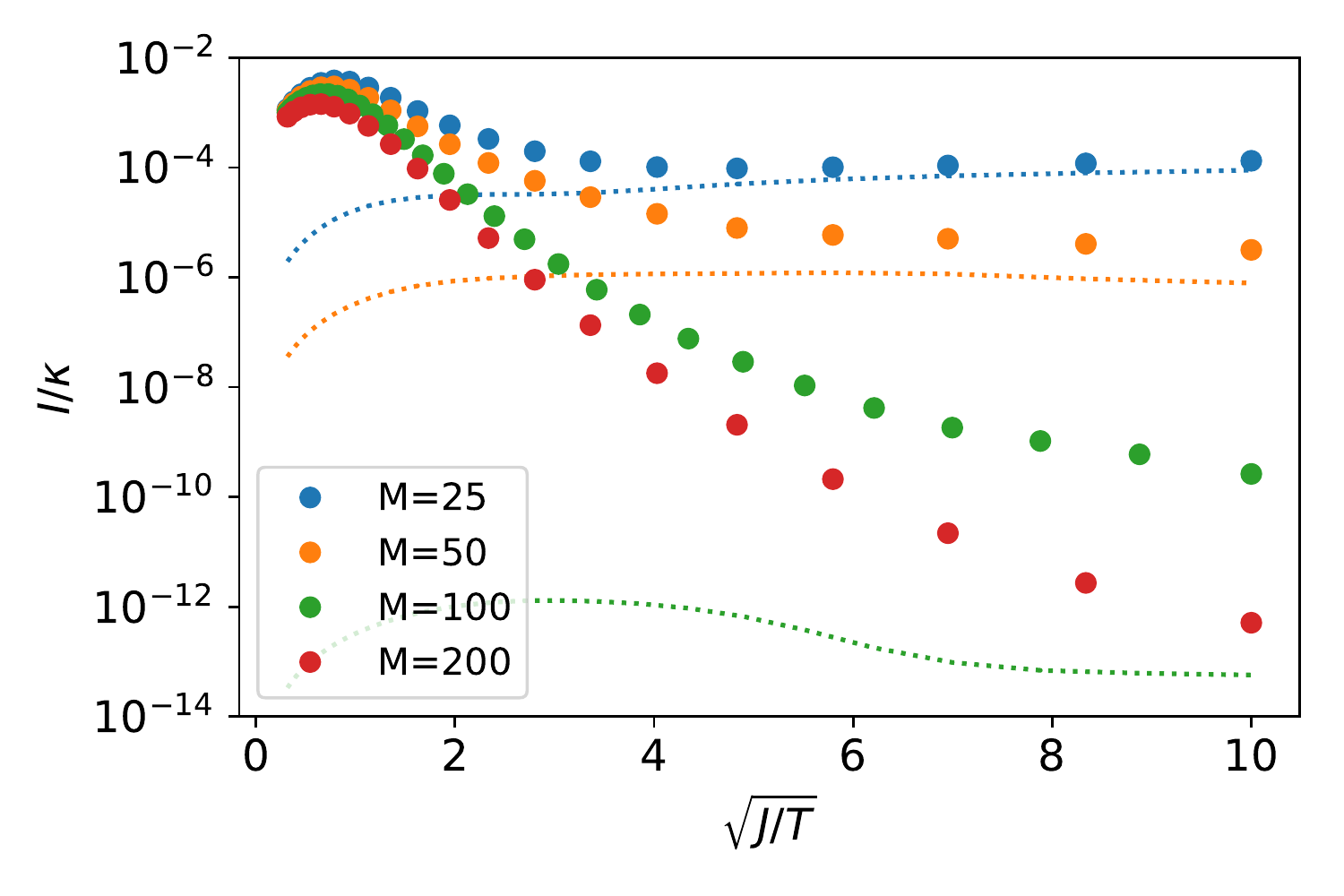}}
    \caption{Current $I$ as a function of the temperature $T$ of the bath for various disorder strengths.
    The dashed lines are fitting curves according to $I = I_s e^{-\sqrt{T_0/T}}$ using the data with $T$ ranging from $0.1J$ to $J$.  The dotted lines are the results in the absence of the global heat-exchange bath [Eq.~(\ref{I0})].
    }\label{T_fit}
\end{figure*}
\begin{figure*}
    \centering
    \subfloat[][$T=0.02J,  W=2J$] {\includegraphics[width=0.5\columnwidth]
      {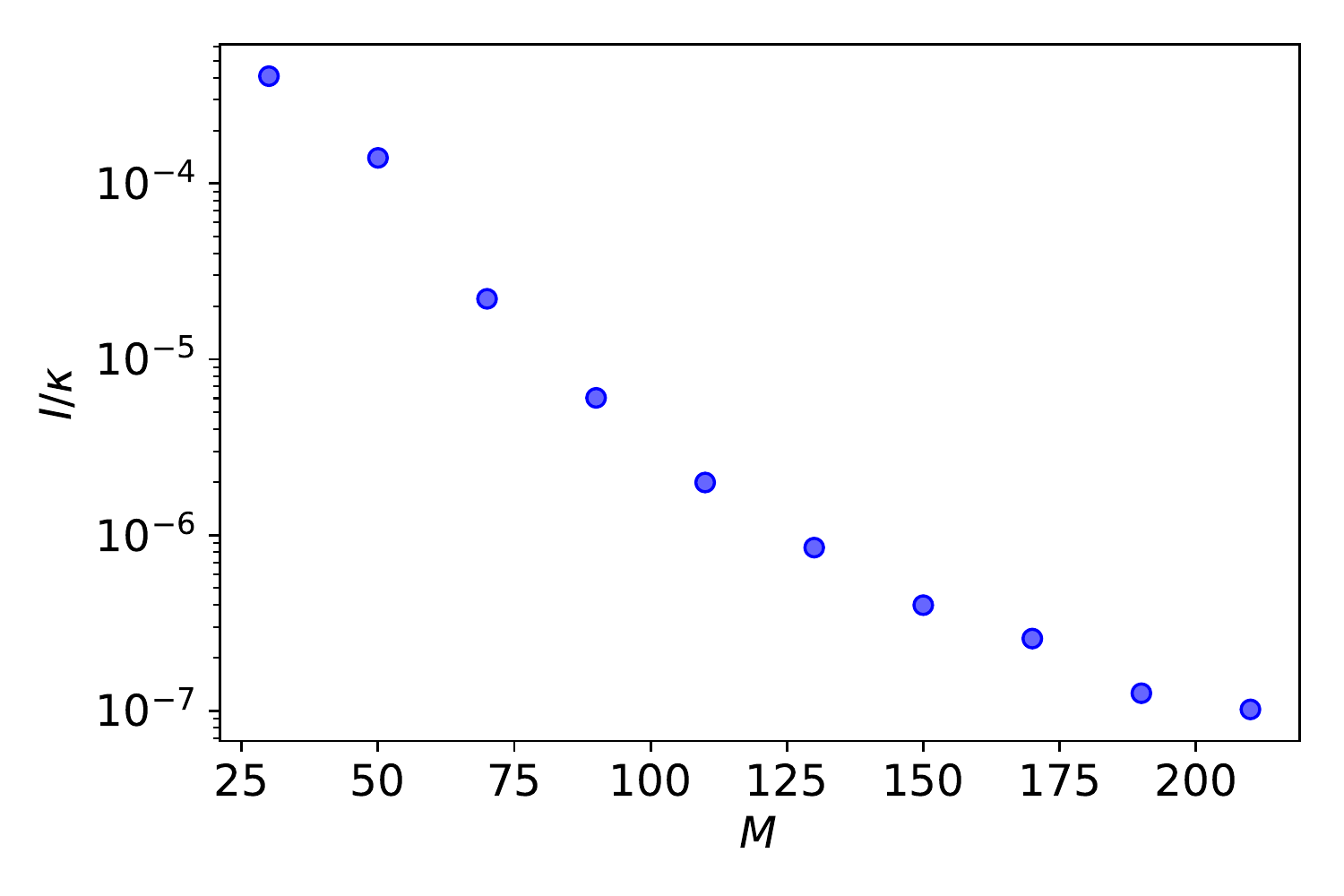}}
    \subfloat[][ ]{\includegraphics[width=0.5\columnwidth]{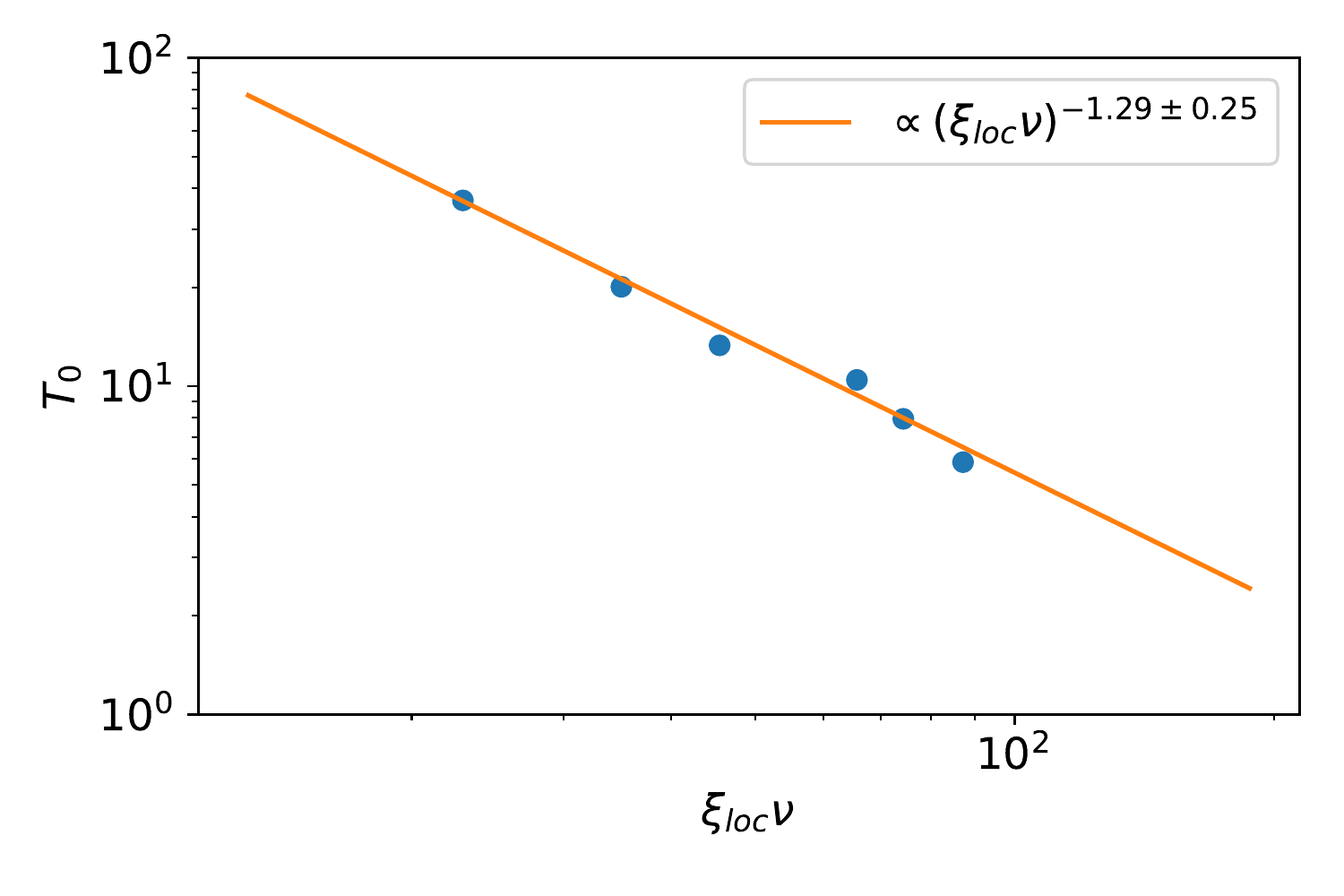}}

    \caption{(a) Current $I$ at low temperature $T=0.02J$ as a function of system size $M$. (b) Mott temperature $T_0$
    (obtained from fitting) as a function of the localization length $\xi_{\text{loc}}$ times the density of states $\nu$ at $k=M/2$.
    }\label{highT}
\end{figure*}
The independence of the current on temperature in the
`low' temperature regime is due to the suppression of heat exchange, which can be inferred from the good agreement
of the results with that without coupling to the global thermal bath  [Eq.~(\ref{I0}), dotted lines].
Nevertheless, the remaining current $(T\rightarrow 0 $) is a finite system size effect~\cite{Marko2016PhysRevLett.117.040601}.
Indeed, as $T\rightarrow 0$, $I_{\gamma = 0}$ decays exponentially with system size $M$, as shown in Fig.~\ref{highT}(a).
For a given system size, $I_{\gamma = 0}$  depends on the disorder strength, and as expected it is smaller for stronger disorder [Fig.~\ref{T_fit}(a)].

The intermediate-temperature VRH regime  can be understood using the following argument due to Mott~\cite{Mott1969}.
Let's consider the hop between states with localization centers separated by the distance $\Delta R$ and with energy difference $\Delta E$.
On the one hand, the probability
to hop is proportional to the envelop overlap between the two states, thus it decays exponentially with the distance $\Delta R$  [$\sim \exp({-2\Delta R/\xi_{\text{loc}}})$].
On the other hand, the probability to
produce excitations of order $\Delta E$ due to the presence of the heat bath is given by the Boltzmann factor ${\rm exp}(-\Delta E/T)$.
This leads
us to assume that the current (conductivity) to leading order is given by
\begin{equation}
I \sim e^{-2\Delta R/\xi_{\text{loc}} -\Delta E/T}.
\end{equation}
As already mentioned, the spatial distance $\Delta R$ and the energy separation $\Delta E$, are not independent, but show clear
anticorrelations (Fig.~\ref{RE}), which shall be captured by $\Delta R \sim (\Delta E \nu)^{-1}$, where  $\nu$ is the density of states. Thus,
the current is determined by the competition between the overlap term ${\rm  exp}(-2\Delta R/\xi_{\text{loc}})$ which favors
short hops and the energy activation $\exp({-\Delta E/T})$, which favors long hops.
Maximizing this probability over $\Delta E$, one finds in one spatial dimension
\begin{equation}
I \propto e^{-2\Delta R_0/\xi_{\text{loc}}} = e^{-\sqrt{T_0/T}} = e^{-\Delta E_0/T},
\end{equation}
 with Mott's hopping length $\Delta R_0 = \sqrt{\xi_{\text{loc}}/(2T\nu)}$ and Mott's temperature $T_0 = {2}/({\xi_{\text{loc}}\nu})$.
Moreover, Mott's hopping energy $\Delta E_0= \sqrt{T_0 T}$ is the energy scale that defines the width of the energy interval of the activated eigenenergies.

 \begin{figure*}
    \centering
\subfloat[]{{\includegraphics[width=0.5\columnwidth]
{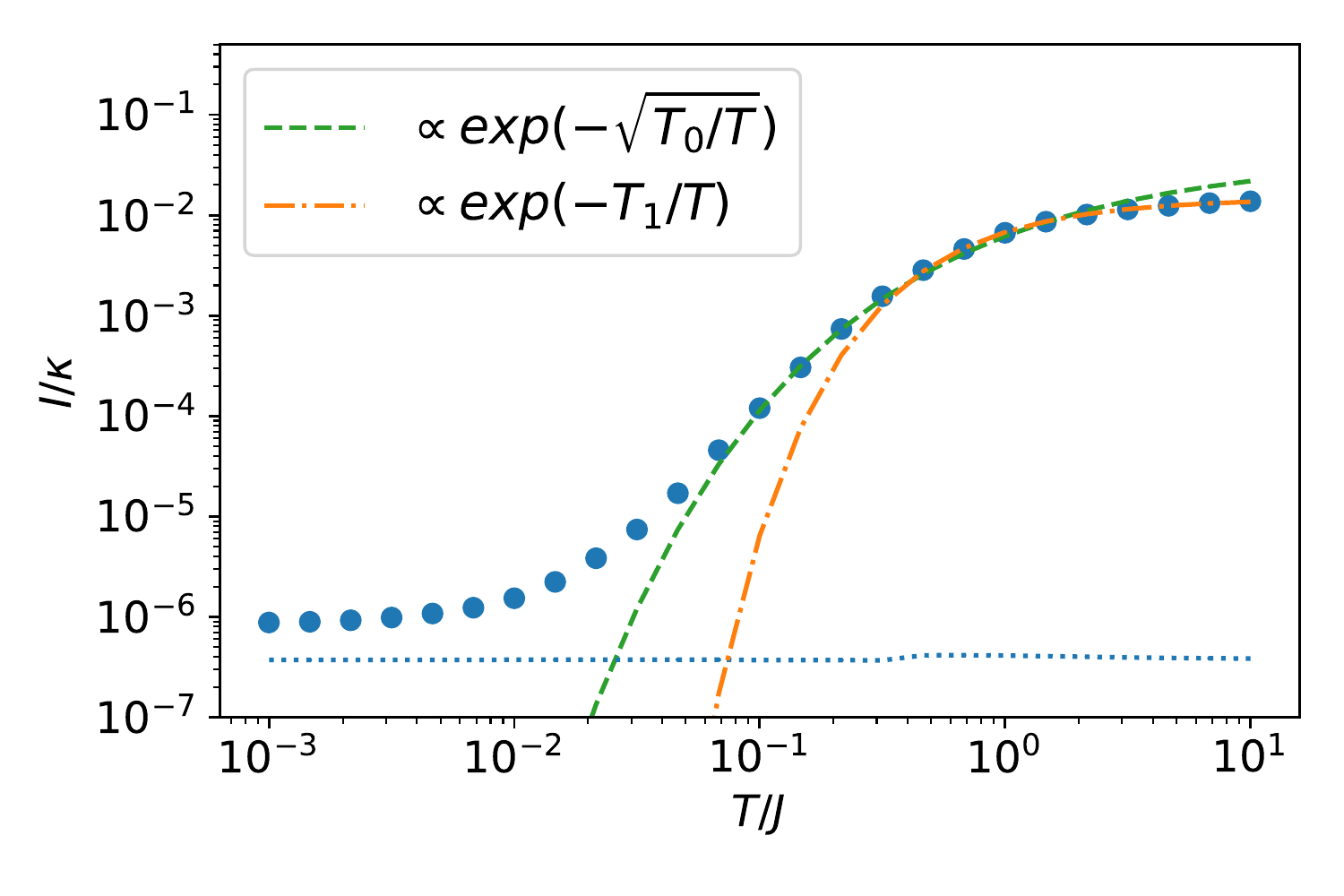}}}
\subfloat[]{\includegraphics[width=0.5\columnwidth]{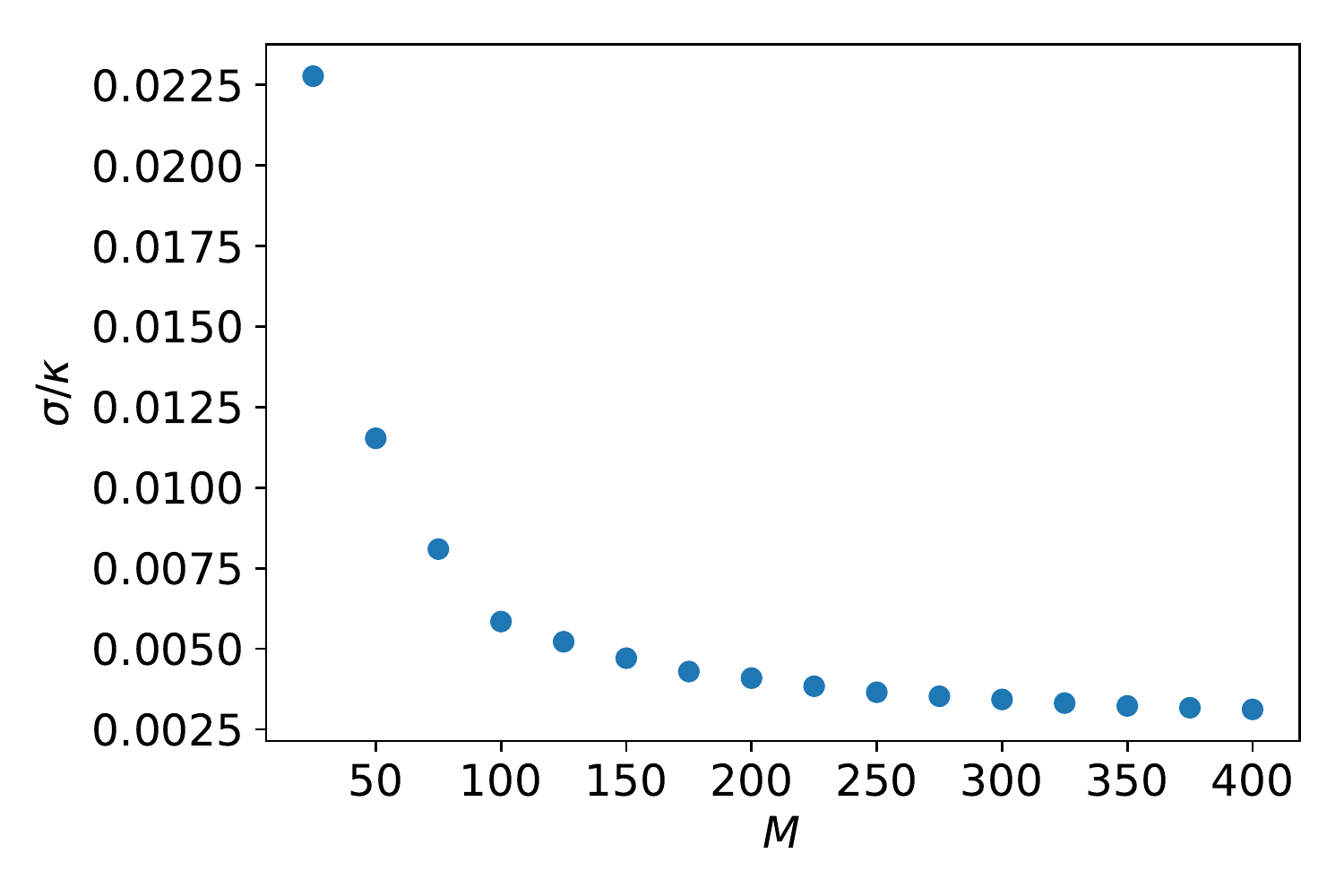}}
    \caption{(a) Current $I$ as a function of the temperature $T$ for the global bath.  Here, the temperatures for the local baths are fixed at $0.01J$. The dotted line is the result in the absence of the global thermal bath, i.e., Eq. (\ref{I0}).
   The disorder strength is $W=2J$. (b) Conductivity $\sigma$ as a function of system size $M$.  Other parameters are $T=0.5J$,
   disorder strength $W=5J$.
    }\label{T1}\label{sigma}
\end{figure*}
By fitting the current in the intermediate temperature regime to Mott's law $I = I_s \exp({-\sqrt{T_0/T}})$, we can extract $T_0$.
Figure~\ref{highT}(b) shows $T_0$  as a function of $\xi_{\text{loc}}\nu$,
where  $\xi_\text{loc}$ is the single-particle localization length in the middle of the energy-band $(\varepsilon_k \sim 0)$.
We find that at strong disorder, $T_0\propto (\xi_{\text{loc}} \nu)^{-1.29\pm 0.25}$,
agreeing rather well with the Mott's prediction $\propto (\xi_{\text{loc}} \nu)^{-1}$.

Finally, in the `high' temperature regime, we attribute the decrease of the current with respect to temperature to the fact that
the difference between
the Fermi distributions of the leads ($f_L(\langle \hat{\tilde{\varepsilon}}_l \rangle)- f_R(\langle \hat{\tilde{\varepsilon}}_l \rangle)$) is
washed out.
To support this idea, we fix the temperature of the leads at $T_L = T_R = 0.01J$.
In this case, the current $I$ increases with the temperature of the global thermal bath $T$ as $\exp({-{T_1/T}})$ in the `high' temperature regime, as shown in Fig.~\ref{T1}(a).
The reason is that the thermal energy $T$ is so
high that, despite of their large energy separation, already activated hopping between nearest neighboring localized orbitals
dominates over the variable-range hopping behavior.


From the current, we can deduce the conductance $G= I/\delta\mu$, 
as well as the conductivity
\begin{equation}
\sigma = GM = \frac{IM}{\delta\mu}.
\end{equation}
Figure \ref{sigma}(b) shows the conductivity $\sigma$ at temperature $T=0.5J$ as a function of the system size $M$
for $W=5J$. We can observe that $\sigma$
converges to a finite value in the limit  $M\to \infty$.

\section{Results II: Variable-range hopping for interacting Fermions}

Let us now discuss transport in the presence of weak interactions, for which the system is in the MBL phase.
Here, we  exploit the advantages of our method, allowing for the treatment of interacting systems in the presence of a thermal environment.
All of our results are obtained in the limit of weak interactions and strong disorder 
in which the approximation [Eq.~(\ref{Happ})] is justified.

\subsection{Transport as a function of temperature $T$}
\begin{figure*}
    \centering
    \subfloat[][ ]{\includegraphics[width=0.5\columnwidth]{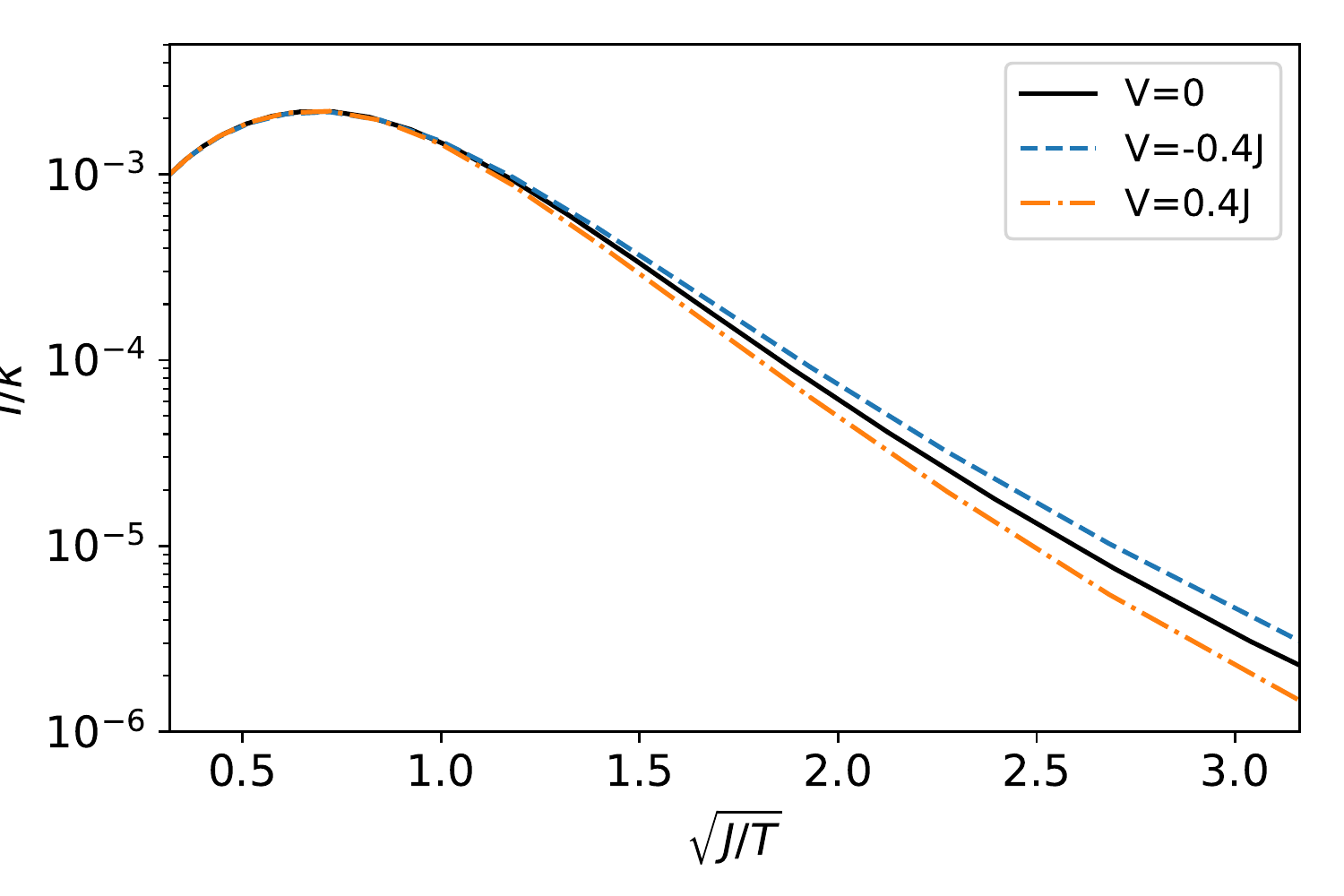}}
     \subfloat[][ ]{\includegraphics[width=0.5\columnwidth]{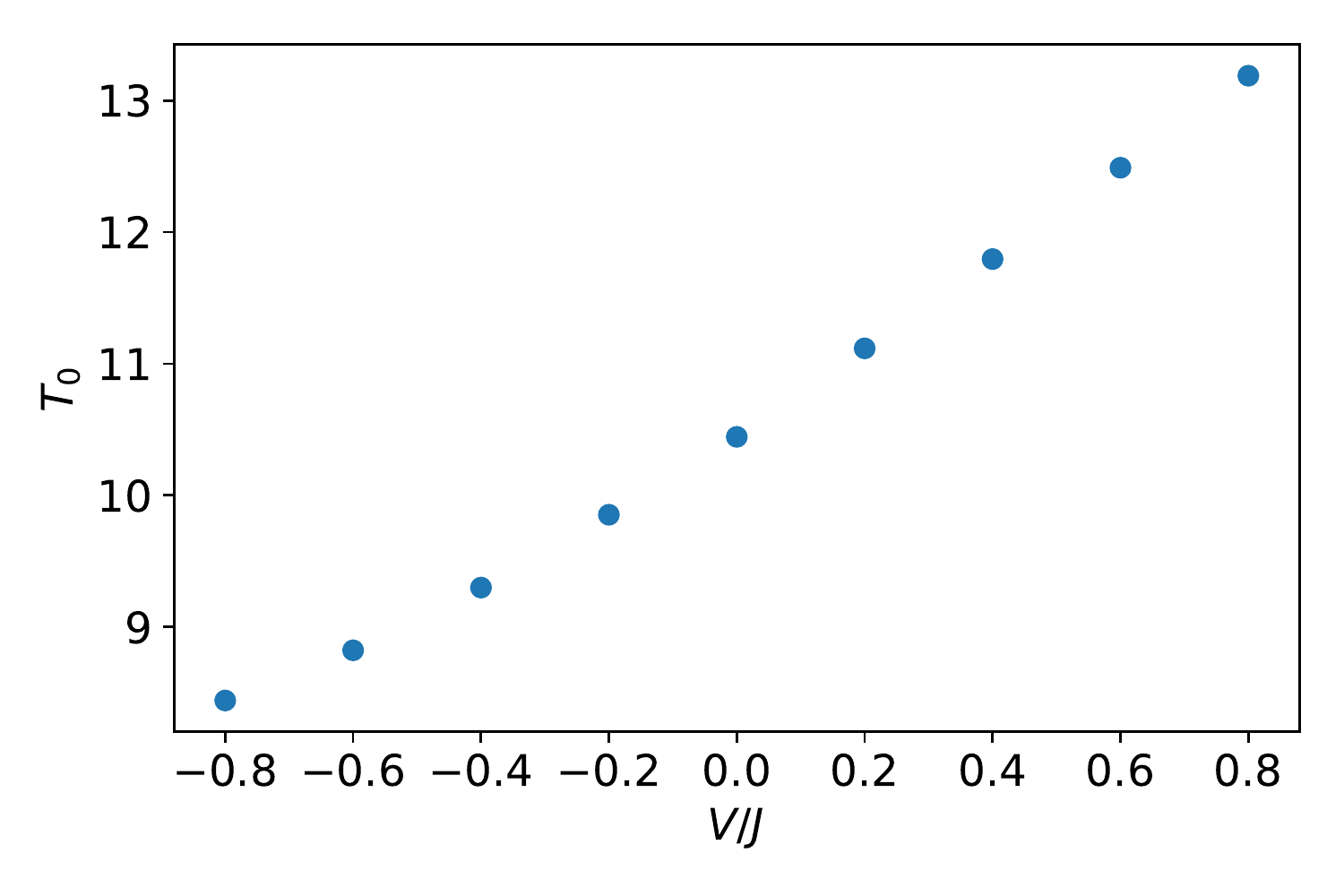}}
    \caption{(a) Temperature dependence of the current for various interaction strengths $V$.  (b) Fitted values of Mott's temperature $T_0$ as a function of the interaction strength $V$.  The disorder strength is $W=3J$.
    }\label{V-obs}
\end{figure*}
Figure~\ref{V-obs}(a) shows the temperature dependence of the current for several interaction strengths $V$.
Interestingly, also for the interacting case $(V\ne 0)$, we find a regime of temperatures,  where the current is explained by Mott's law
($I\sim \exp({-\sqrt{T_0/T}})$).
We find that in this VRH regime, attractive interactions (blue dashed line) enhance transport,
while for repulsive interactions (orange dash-dotted line) the current decreases.
This means that the interaction changes Mott's temperature $T_0$.
Fig.~\ref{V-obs}(b) shows the dependence of Mott's temperature $T_0$ on the interaction strength $V$.
We can see  how attractive interactions decrease $T_0$, while repulsive interactions increase $T_0$.

Can we understand this behavior?
From our previous discussion we know that $T_0$ depends on the localization length $\xi_{\text{loc}}$ and on the density of states $\nu$.
However, a change of the $l$-bits and their localization length is expected to be of second order with respect to interactions (and is also not taken into account in the approximate Hamiltonian~\eqref{Happ}). We should, therefore, be able to explain the interaction-induced shift of Mott's temperature in terms of interaction-induced shifts of the density of states.

\subsection{Interaction-shifted density of states}
\begin{figure*}
    \centering
      \subfloat[][]{\includegraphics[width=0.5\columnwidth]{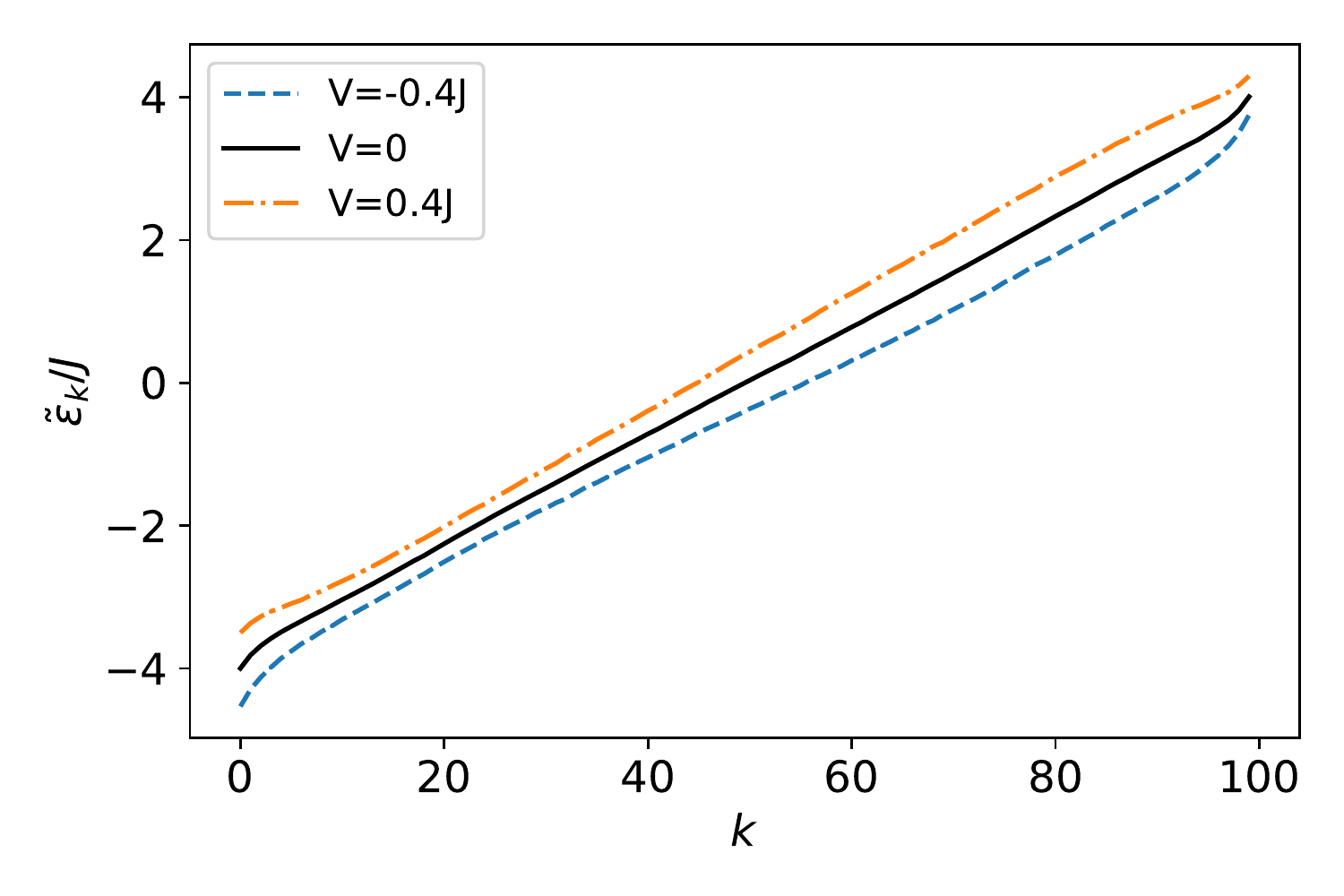}}
       \subfloat[][]{\includegraphics[width=0.5\columnwidth]{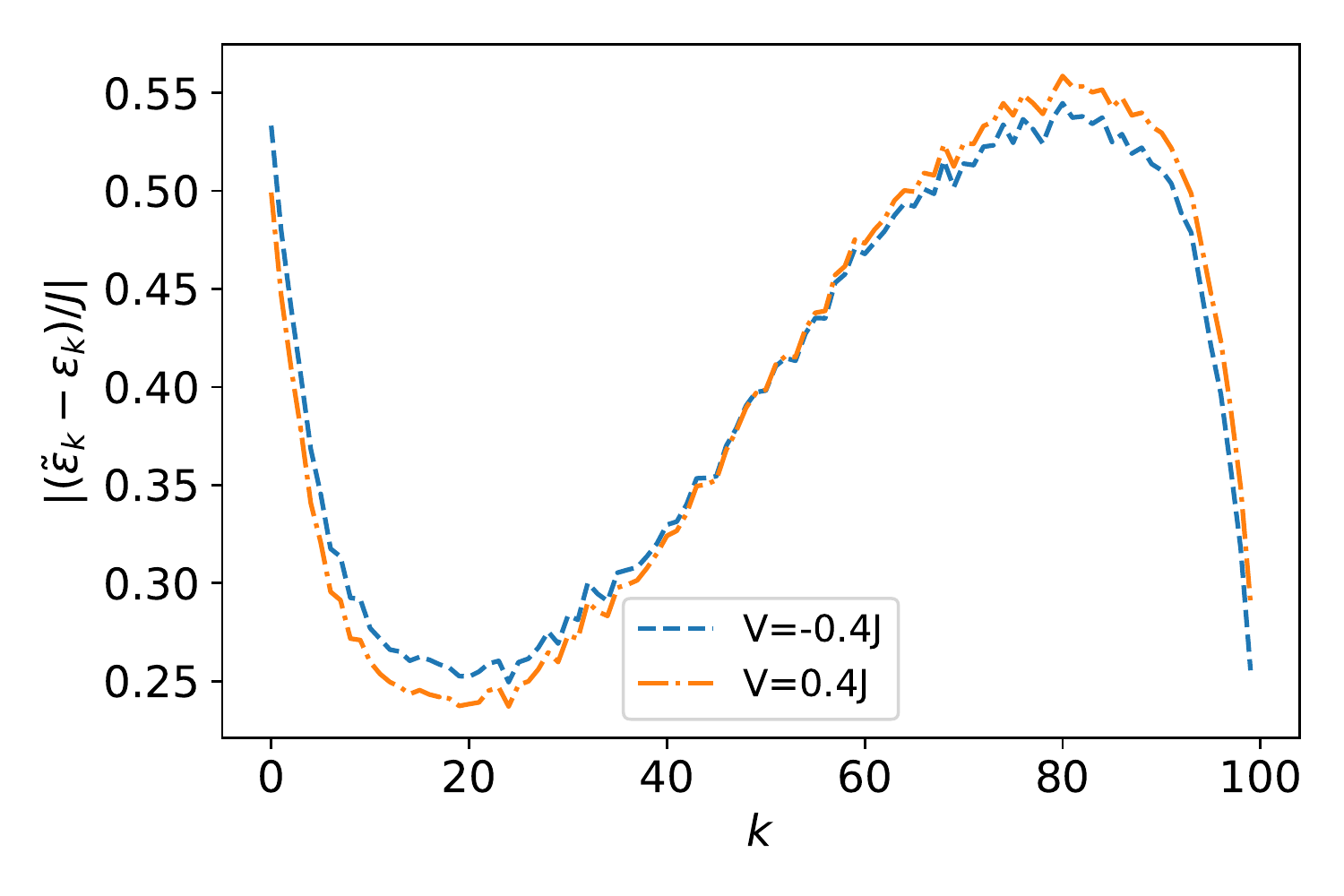}}
    \caption{(a) Interaction-shifted single-particle energies $\langle \hat{\tilde{\varepsilon}}_k \rangle$ at temperature $T=0.5J$. (b) Magnitude of  the energy shift $\delta E_k \equiv \langle \hat{\tilde{\varepsilon}}_k \rangle- {\varepsilon}_k  $.
    }\label{V-expl}
\end{figure*}

{In Fig.~\ref{V-expl}(a), we show the  energies $\langle \hat{\tilde{\varepsilon}}_k \rangle$
in the steady state for $V=-0.4J$, $V=0$ and $V=0.4J$ all at $T=0.5J$.
As expected from Eq.~(\ref{Ek}),  repulsive interactions ($V>0$, orange dash-dotted line) shift the single-particle energies ${\varepsilon}_k$ up
(black solid line) and  attractive interactions ($V<0$, blue dashed line) decrease them.
However, this does not imply a change in the average level splitting $\langle \hat{\tilde{\varepsilon}}_k-\hat{\tilde{\varepsilon}}_{k-1}\rangle$.
What is crucial is rather that the interaction-dependent energy shift depends on $k$.
The  absolute value of the energy shift
$\delta E_k \equiv \langle \hat{\tilde{\varepsilon}}_k \rangle - {\varepsilon}_k$ is shown in Fig.~\ref{V-expl}(b).
For the eigenstates in the middle of the spectrum, which are the ones that contribute
predominantly to the transport,
the energy shift increases with energy.

This behavior can be understood from considering Eq.~(\ref{Ek}):
The energy shift $\delta E_k =  \sum_q {U_{kq} \langle \hat{n}_q \rangle}$
depends via $U_{kq}$ on the overlap of the involved single-particle wave functions. Therefore, the main contribution to the shift originates from eigenstates that are close by in space.
Now, due to the anti-correlation property between energetic and spatial distance, these eigenstates $q$ will have
a large energy difference with respect to $k$.
Thus a state $k$ slightly above the Fermi energy will have more likely neighboring states below the Fermi energy with a large occupation probability, while a state slightly below the Fermi energy will more likely have neighboring states above the Fermi energy, with a small occupation probability. In this way, the positive/negative energy shift of repulsive/attractive interactions, will be larger for states above the Fermi level than for states below it.
This implies that the level spacing between neighboring energy levels $(\langle \hat{\tilde{\varepsilon}}_k-\hat{\tilde{\varepsilon}}_{k-1}\rangle)$ is increased for repulsive interactions
and is decreased for attractive interactions. In other words, repulsive interactions reduce the density of states $\nu_k = 1/(\langle\hat{\tilde{\varepsilon}}_k-\hat{\tilde{\varepsilon}}_{k-1}\rangle)$
and attractive interactions enhance it.

\subsection{Explanation of the change of Mott's temperature}
\begin{figure*}
    \centering
     \subfloat[][]{\includegraphics[width=0.5\columnwidth]{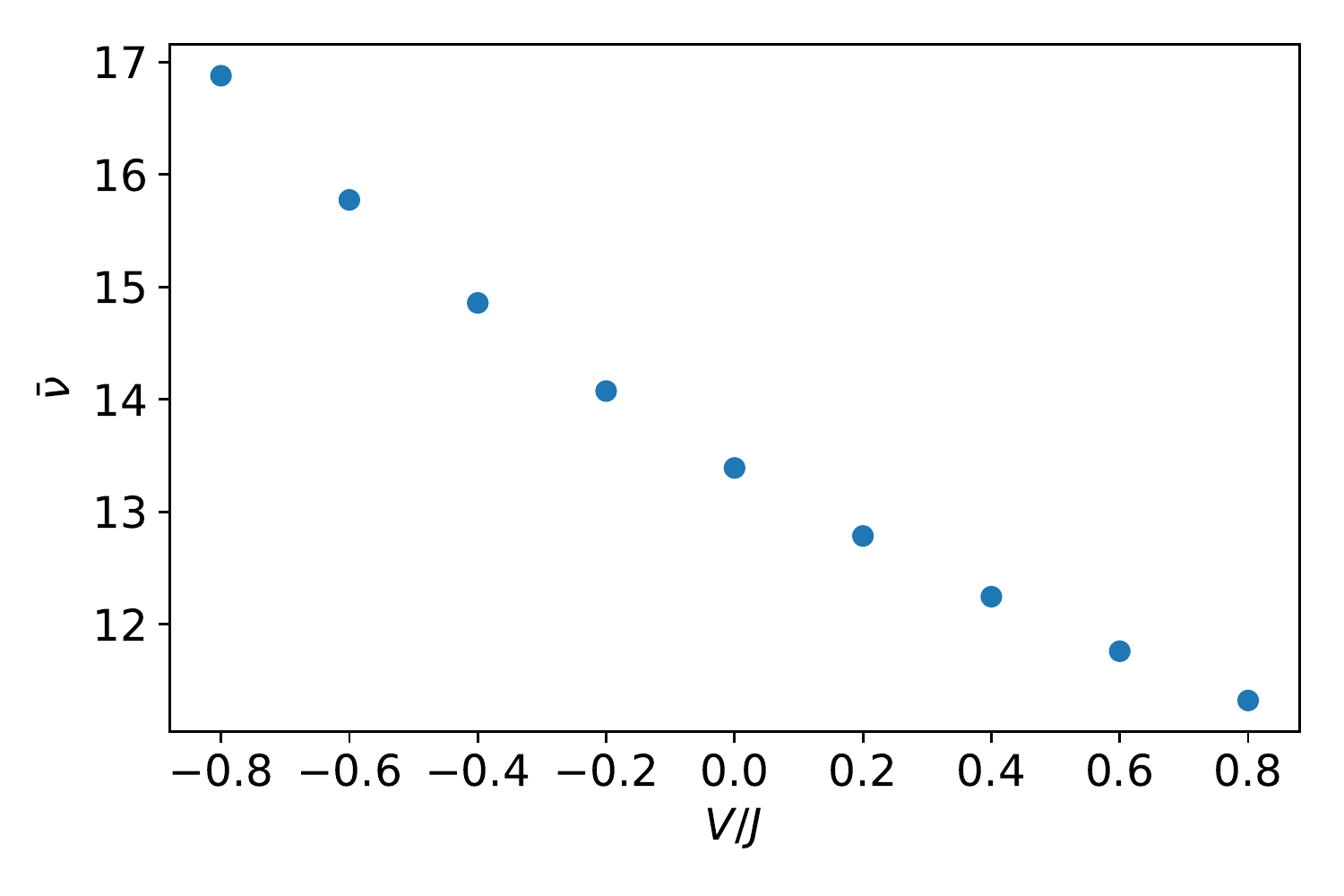}}
       \subfloat[][]{\includegraphics[width=0.5\columnwidth]{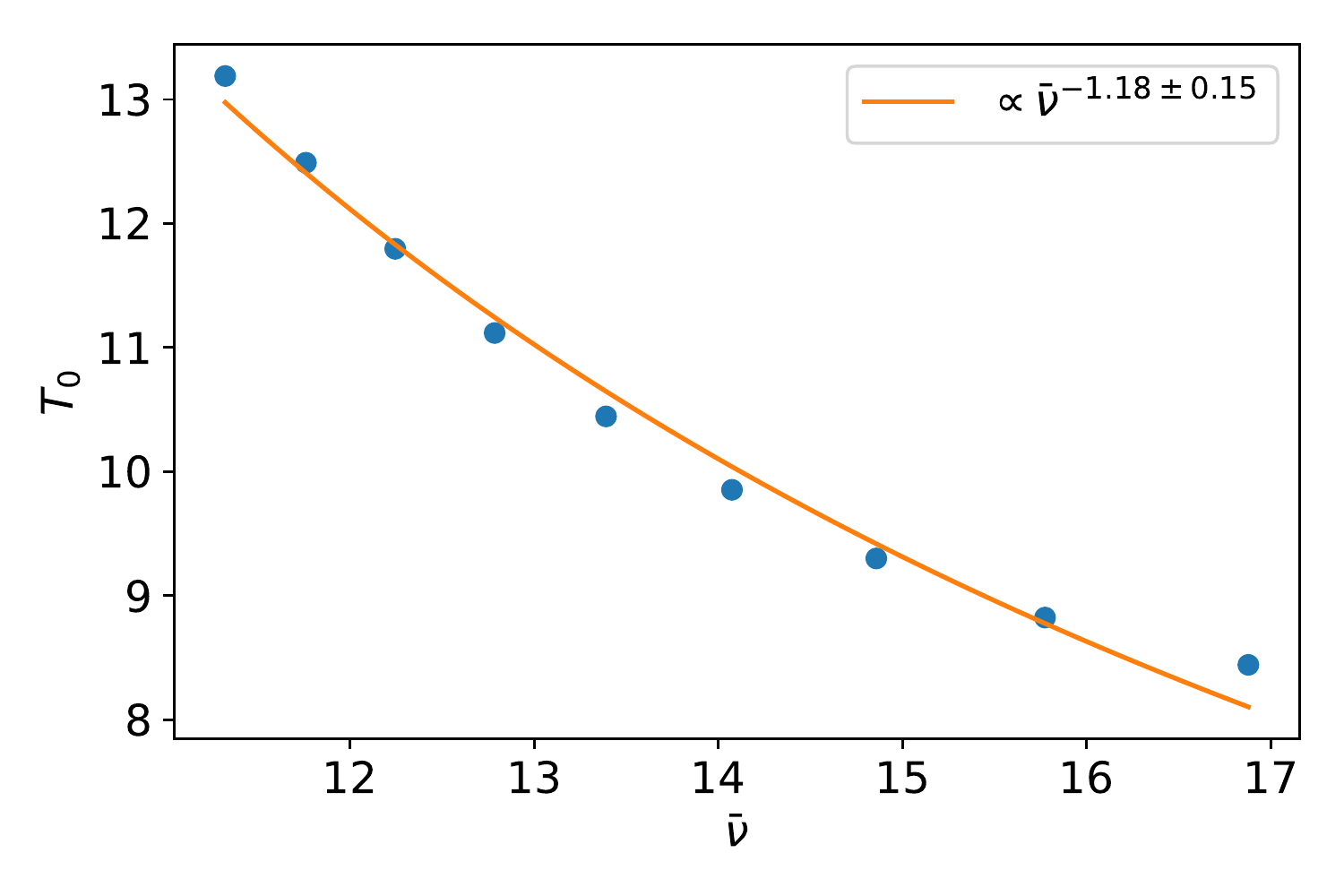}}
    \caption{(a) Averaged density of states (over states with $k$ between $40$ and $60$) $\bar \nu$ as a function of the interaction strength at temperature $T=0.5J$. (b) Fitted values of Mott's temperature $T_0$ as a function of the averaged density of states $\bar \nu$.
    }\label{V}
\end{figure*}

The suspected behavior is confirmed in Fig.~\ref{V}(a), where we show the averaged density of states
 (over states with $k$ between $40$ and $60$), $\bar \nu $,
as a function of interaction strength at temperature $T=0.5J$.
We can now check whether the dependence of Mott's temperature $T_0$ on interactions can be explained
with their effect on the density of states. When we plot $T_0$
 versus $\bar{\nu}$, as shown in Fig.~\ref{V}(b), we observe a behavior $T_0 \propto \bar{\nu}^{-1.18 \pm 0.15}$, which agrees well with the predicted
$\nu^{-1}$ dependency in Mott's temperature.

%

%
We would like to emphasize that the origin of Mott's law we find is different from the case of Coulomb interactions,
 which has been shown to lead to a variable-range hopping conductivity $\sigma \propto \exp[-(T_0/T)^{1/2}]$ independent of dimensionality due to a nonanalytic modification of the density of states near the Fermi energy~\cite{1975JPhC....8L..49E}.
In our case of local interactions of nearest-neighbor type the density of states becomes modified only smoothly which continuously connects the interacting with the noninteracting result.

 }
\section{Conclusion}

In this work we have introduced a method which allows for an efficient description of MBL systems at strong disorder and weak interactions, when weakly coupled to thermal environments.
We have benchmarked our method for noninteracting systems, where we showed that our method recovers Mott's law from variable-range hopping starting from a microscopic model.
Upon adding weak interactions we have found that Mott's law persists while leading to perturbative corrections.
We explain our observations by an interaction-induced modification of the density of states due to spatio-energetic correlations.
Already in recent years the study of MBL systems coupled to environments has received substantial interest not only in  experiments~\cite{Bloch2017PhysRevX.7.011034,Rubio2018arXiv}.
On the theoretical side these previous studies~\cite{Levi2016PhysRevLett.116.237203,Fischer2016PhysRevLett.116.160401,Medvedyeva2016PhysRevB.93.094205,2017PhRvB..95s5135L,2018arXiv180604772L,Marko2016AP,Everest2017PhysRevB.95.024310,Nandkishore2016AP,Gopalakrishnan2017PhysRevLett.119.046601,Nandkishore2014PhysRevB.90.064203,Johri2015PhysRevLett.114.117401,Nandkishore2015PhysRevB.92.245141,Luitz2017PhysRevLett.119.150602,Hyatt2017PhysRevB.95.035132,Yusipov2017PhysRevLett.118.070402} have been either limited in the accessible system sizes~\cite{Levi2016PhysRevLett.116.237203,Medvedyeva2016PhysRevB.93.094205,2017PhRvB..95s5135L,2018arXiv180604772L} or make rather specific assumptions regarding the properties of the environment, e.g. it is described by dephasing noise corresponding to an infinite temperature environment~\cite{Levi2016PhysRevLett.116.237203,Fischer2016PhysRevLett.116.160401,Marko2016AP,Medvedyeva2016PhysRevB.93.094205,Everest2017PhysRevB.95.024310}, that it is described by classical noise~\cite{Nandkishore2016AP,Gopalakrishnan2017PhysRevLett.119.046601}, or that it is given by a small bath of size comparable to the system~\cite{Nandkishore2014PhysRevB.90.064203,Johri2015PhysRevLett.114.117401,Nandkishore2015PhysRevB.92.245141,Luitz2017PhysRevLett.119.150602,Hyatt2017PhysRevB.95.035132}.
In the present work we provided an alternative approach, which allows for an efficient solution for mesoscopic system sizes in contact with conventional thermal environments.
This might also be of particular importance in view of experiments in quantum simulators, where the influence of environments has been studied systematically recently~\cite{Bloch2017PhysRevX.7.011034,Rubio2018arXiv}.
Within the presented formulation the method is limited to MBL systems at strong disorder and weak interactions.
However, this limitation originates solely from the method used to diagonalize the system Hamiltonian following the recent work~\cite{Giu18}.
This can be improved upon finding a more accurate diagonalization of the MBL Hamiltonian, i.e., a more accurate $l$-bit description, following, e.g., some recent ideas~\cite{2016PhRvB..94d1116P,2017PhRvX...7b1018W}.
This would not influence substantially the structure of the quantum master equation, but would allow to study the steady states in broader parameter range, e.g., also closer to the MBL transition towards ergodicity.
Concluding, our work provides a framework to study open system dynamics at mesoscopic scales for various scenarios involving MBL systems at strong disorder and not too strong interactions.
This includes a wide range of phenomena such as MBL-spin glasses, MBL topological phases, or time crystals.

\section*{Acknowledgements}
We acknowledge support from the Deutsche Forschungsgemeinschaft (DFG) via the Research Unit FOR 2414 (grant number EC 392/3-1) and via the Gottfried  Wilhelm  Leibniz  Prize  program.

\section{Appendix}

\subsection{Quantum-Jump Monte Carlo Simulation}\label{QJMC}
We use the Gillespie algorithm~\cite{DanielPRE} to perform the time evolution. For each trajectory,
the system is initially prepared in a random state. Then the algorithm alternates between the following two steps. (i) Stay in a Fock state for some time. (ii) Jump to another occupation basis state.
The time interval $\tau$ for staying in the current state is determined by $\tau = {\rm min}(\tau_h, \tau_g,\tau_l)$, the minimum of three values randomly drawn from exponential distribution $P(\tau_{\lambda}) \propto {\rm exp}[-\tau_{\lambda}/\bar t_{\lambda}]$ with mean dwell time for heat exchange ($\bar t_h $), gaining ($\bar t_g$) and losing ($\bar t_l $) a particle given by
\begin{align}
\bar t_h&= \sum\limits_{k,q}{{R_{{{\bf n}_{qk}},{\bf n}}}}, \notag\\
\bar t_g&=\sum\limits_{k}{ \sum\limits_{\alpha=L,R}{\eta_\alpha(k)f_\alpha(\tilde{\varepsilon}_k)}} , \notag\\
\bar t_l&=\sum\limits_{k}{ \sum\limits_{\alpha=L,R}{\eta_\alpha(k)(1-f_\alpha(\tilde{\varepsilon}_k))}}.
\end{align}
According to the choice made, the corresponding jump operation is  performed. Specifically, if $\tau = \tau_h$, which means heat exchange process is chosen, then a particle is transferred from a randomly drawn departure mode $k$ to the randomly drawn target mode $q$. This single-particle jump has the probability $P(k \rightarrow q, {\bf n}) = \bar t_h R_{{{\bf n}_{qk}},{\bf n}}$. If pumping is chosen, with  $\tau = \tau_g$, a particle is added to a  mode randomly drawn with probability $P(k) = \bar t_p{ \sum_{\alpha=L,R}{\eta_\alpha(k)f_\alpha(\tilde{\varepsilon}_k)}} $. Likewise, the particle loss process is performed when  $\tau = \tau_l$. These two steps are repeated until the desired evolution time is reached.

An ensemble of trajectories is calculated individually, from which
the wavefunctions $| {\bf n}^{(\lambda)}(t)\rangle$ obtained are then used to compute the expectation value of an observable $O$  as
\begin{equation}
\langle O(t) \rangle = \frac{1}{L}\sum\limits_{\lambda=1}^{L}{\langle {\bf n}^{(\lambda)}(t)|O| {\bf n}^{(\lambda)}(t)\rangle}.
\end{equation}

\subsection{Derivation of Eq. (\ref{dnhe}) and (\ref{dnpe}) in the main text}\label{derivation}
The time evolution of the mean occupation due to heat exchange is governed by
\begin{equation}\label{n_heat}
\left(\frac{d}{{dt}}{\langle {{\hat n}_l}\rangle}\right) _{{\rm{heat}}} = \frac{1}{2}\sum\limits_{{\bf n},k,q} {{R_{{{\bf n}_{qk}},{\bf n}}}{\rm tr}\{[2\hat{L}_{qk}^\dag({\bf {n}}) \hat{n}_l \hat{L}_{qk}({\bf {n}}) - \hat{n}_l\hat{L}_{qk}^\dag({\bf {n}})  \hat{L}_{qk}({\bf {n}})-\hat{L}_{qk}^\dag({\bf {n}})  \hat{L}_{qk}({\bf {n}})  \hat{n}_l]\rho\}},
\end{equation}
with
\begin{eqnarray}\label{com}
&& 2\hat{L}_{qk}^\dag({\bf {n}}) \hat{n}_l \hat{L}_{qk}({\bf {n}}) - \hat{n}_l\hat{L}_{qk}^\dag({\bf {n}})  \hat{L}_{qk}({\bf {n}}) -\hat{L}_{qk}^\dag({\bf {n}})  \hat{L}_{qk}({\bf {n}})  \hat{n}_l \notag\\
&=& 2|{\bf{n}}\rangle \langle {\bf{n}}_{qk}|\hat{n}_l|{\bf{n}}_{qk}\rangle \langle{\bf{n}}| - \hat{n}_l|{\bf{n}}\rangle \langle{\bf{n}}| - |{\bf{n}}\rangle \langle{\bf{n}}| \hat{n}_l
\end{eqnarray}
Let us assume that Fock state $|{\bf n }\rangle $ has $n_l$ particle in $l$ mode, i.e.,
\begin{equation}\label{nl_n}
\hat n_l |{\bf n }\rangle = n_l |{\bf n }\rangle.
\end{equation}
 Then it is easy to verify that for state $|{\bf n }_{qk}\rangle $, which is obtained from state $|{\bf n}\rangle$ by transferring a particle from $k$ mode to $q$ mode, there is the following property
\begin{equation}\label{nl_nkq}
\hat n_l |{\bf n }_{qk}\rangle = (n_l+\delta_{q,l}-\delta_{k,l}) |{\bf n }_{qk}\rangle.
\end{equation}
By using Eqs. (\ref{nl_n}) and (\ref{nl_nkq}), we can reduce Eq.~(\ref{com}) to
\begin{equation}\label{com1}
2\hat{L}_{qk}^\dag({\bf {n}}) \hat{n}_l \hat{L}_{qk}({\bf {n}}) - \hat{n}_l\hat{L}_{qk}^\dag({\bf {n}})  \hat{L}_{qk}({\bf {n}}) -\hat{L}_{qk}^\dag({\bf {n}})  \hat{L}_{qk}({\bf {n}})  \hat{n}_l=  2(\delta_{q,l}-\delta_{k,l}) |{\bf{n}}\rangle \langle{\bf{n}}|
\end{equation}
Substituting it into Eq.~(\ref{n_heat}), we obtain
\begin{equation}\label{a6}
\left(\frac{d}{{dt}}\langle {\hat n_l}\rangle\right) _{{\rm{heat}}}= \sum\limits_{{\bf n},k,q} {{R_{{\bf n}_{qk},{{\bf n}}}}(\delta_{q,l}-\delta_{k,l}) p_{{\bf n}}}
=\sum\limits_{{\bf n},k} {\left(p_{{\bf n}}{R_{{\bf n}_{lk},{{\bf n}}}} - p_{{\bf n}}{R_{{\bf n}_{kl},{{\bf n}}}}\right)} .
\end{equation}
It then reduces to Eq.~(\ref {dnhe}) by using Eq.~(\ref{Rn}) and (\ref{oRqk}) in the main text.

Likewise, by making use of
\begin{align}
2\hat{L}_k \hat{n}_l \hat{L}_k^\dag -  \hat{n}_l \hat{L}_k\hat{L}_k^\dag -  \hat{L}_k\hat{L}_k^\dag  \hat{n}_l &= 2\delta_{k,l} |{\bf{n}}_{k\downarrow}\rangle \langle{\bf{n}}_{k\downarrow}| , \notag\\
2\hat{L}_k^\dag  \hat{n}_l \hat{L}_k-  \hat{n}_l \hat{L}_k^\dag\hat{L}_k -  \hat{L}_k^\dag \hat{L}_k \hat{n}_l &= -2\delta_{k,l} |{\bf{n}}\rangle \langle{\bf{n}}|,
\end{align}
we can obtain Eq.~(\ref{dnpe}) in the main text.

\subsection{Anderson localization $V=0$}\label{AppAnderson}
\begin{figure*}[!htbp]
    \centering
    \subfloat[][$W =1 J$]{\includegraphics[width=0.33\columnwidth]{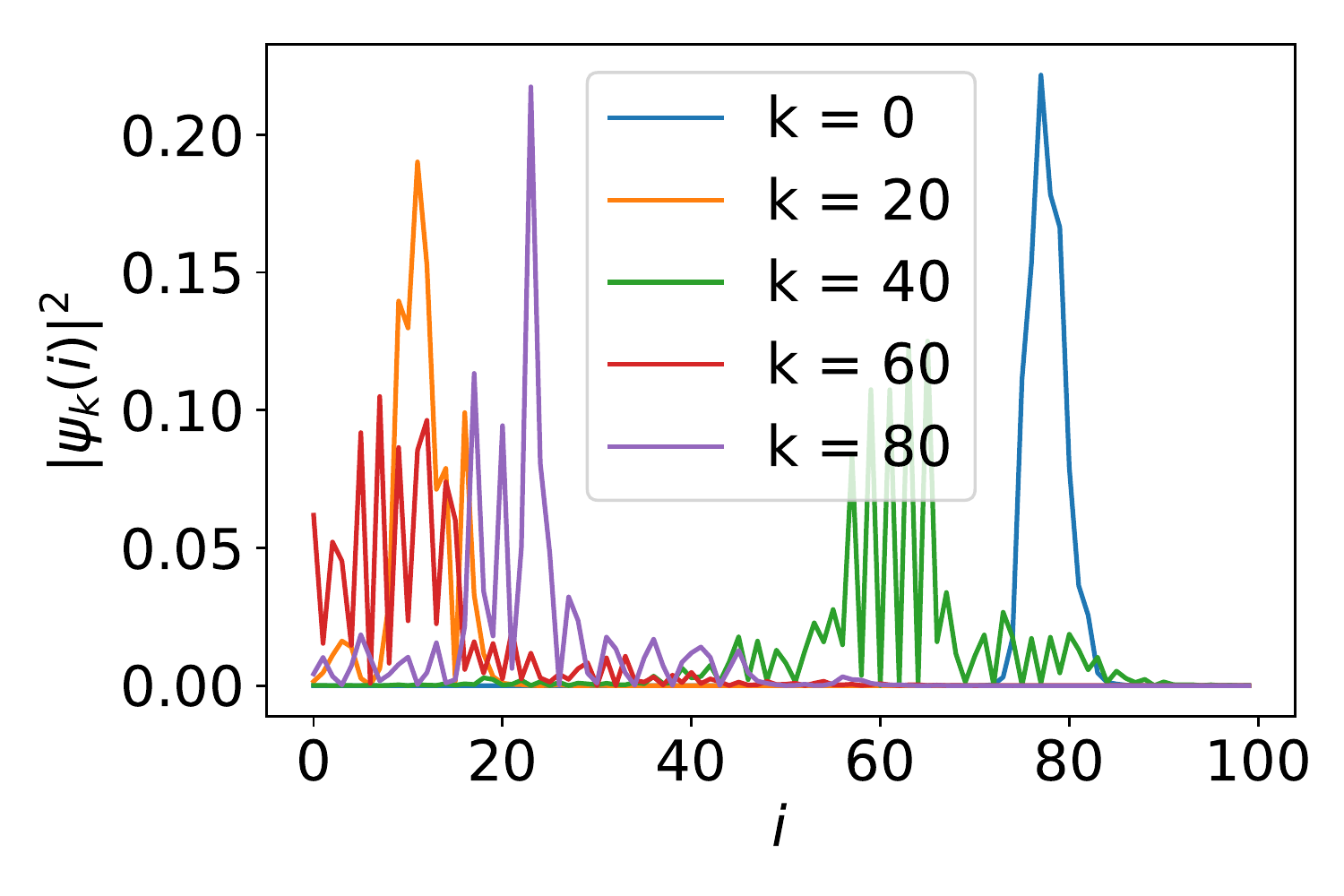}}
     \subfloat[][$W =10 J$]{\includegraphics[width=0.33\columnwidth]{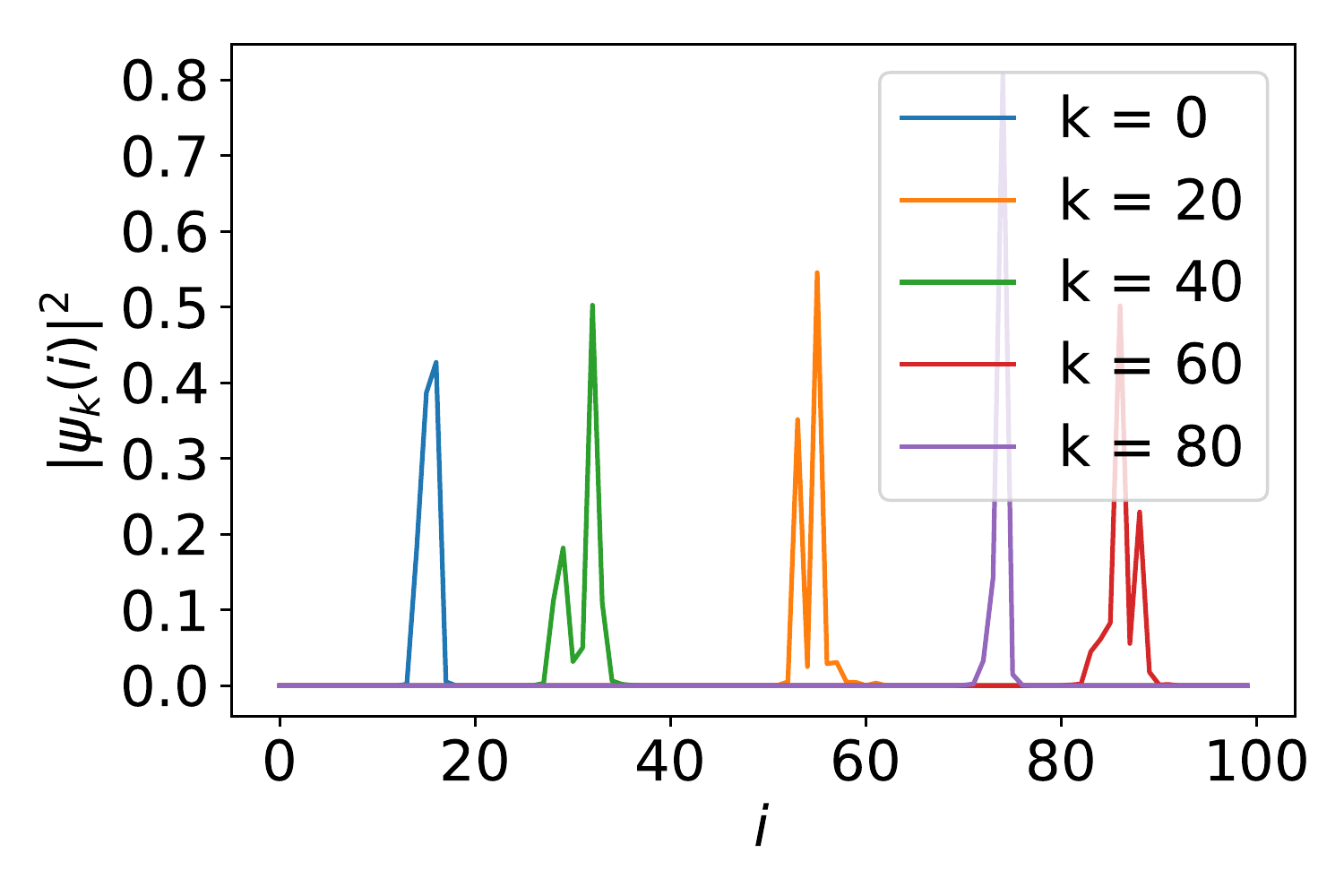}}
      \subfloat[]{\includegraphics[width=0.33\columnwidth]{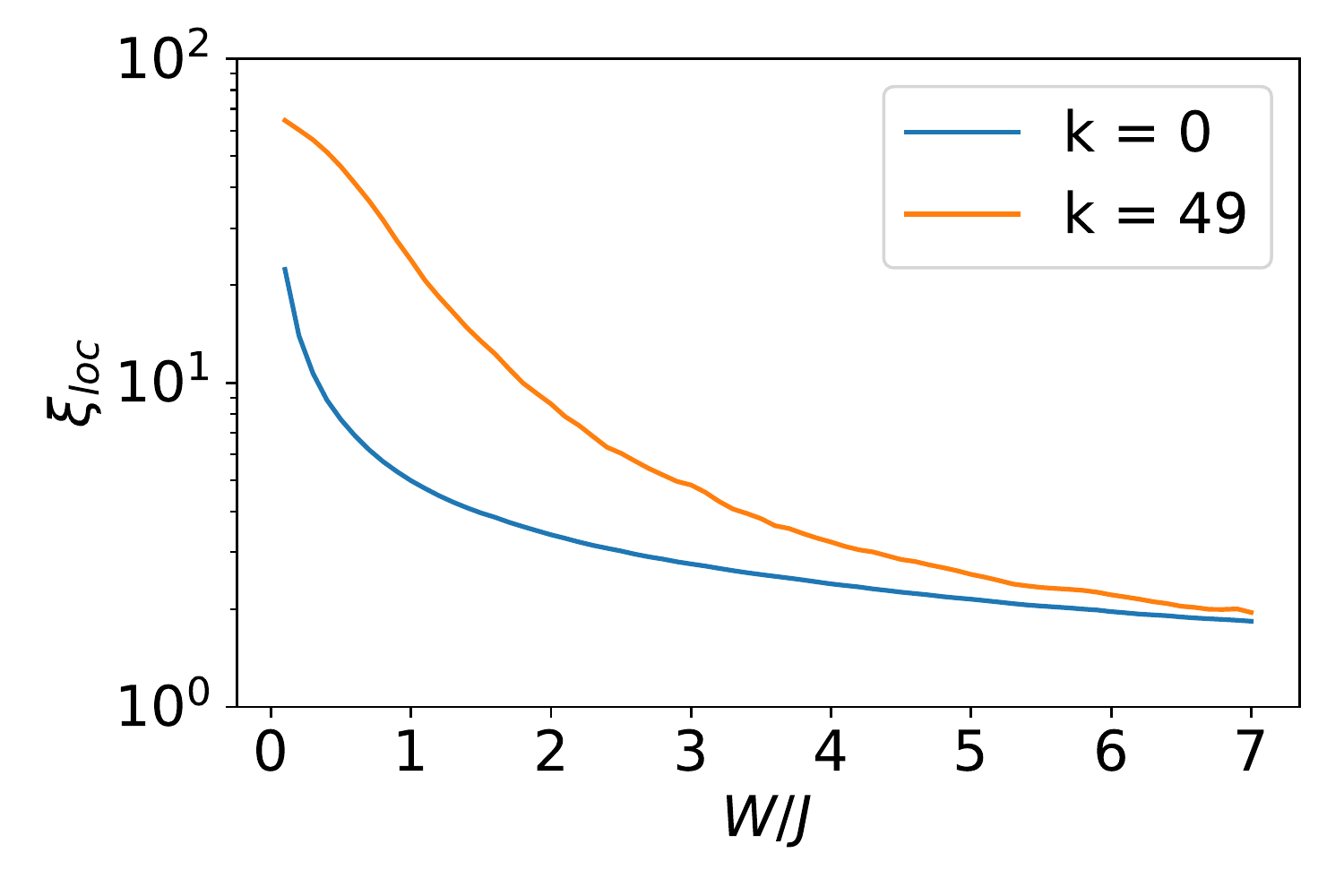}}\\
    \subfloat[]{\includegraphics[width=0.33\columnwidth]{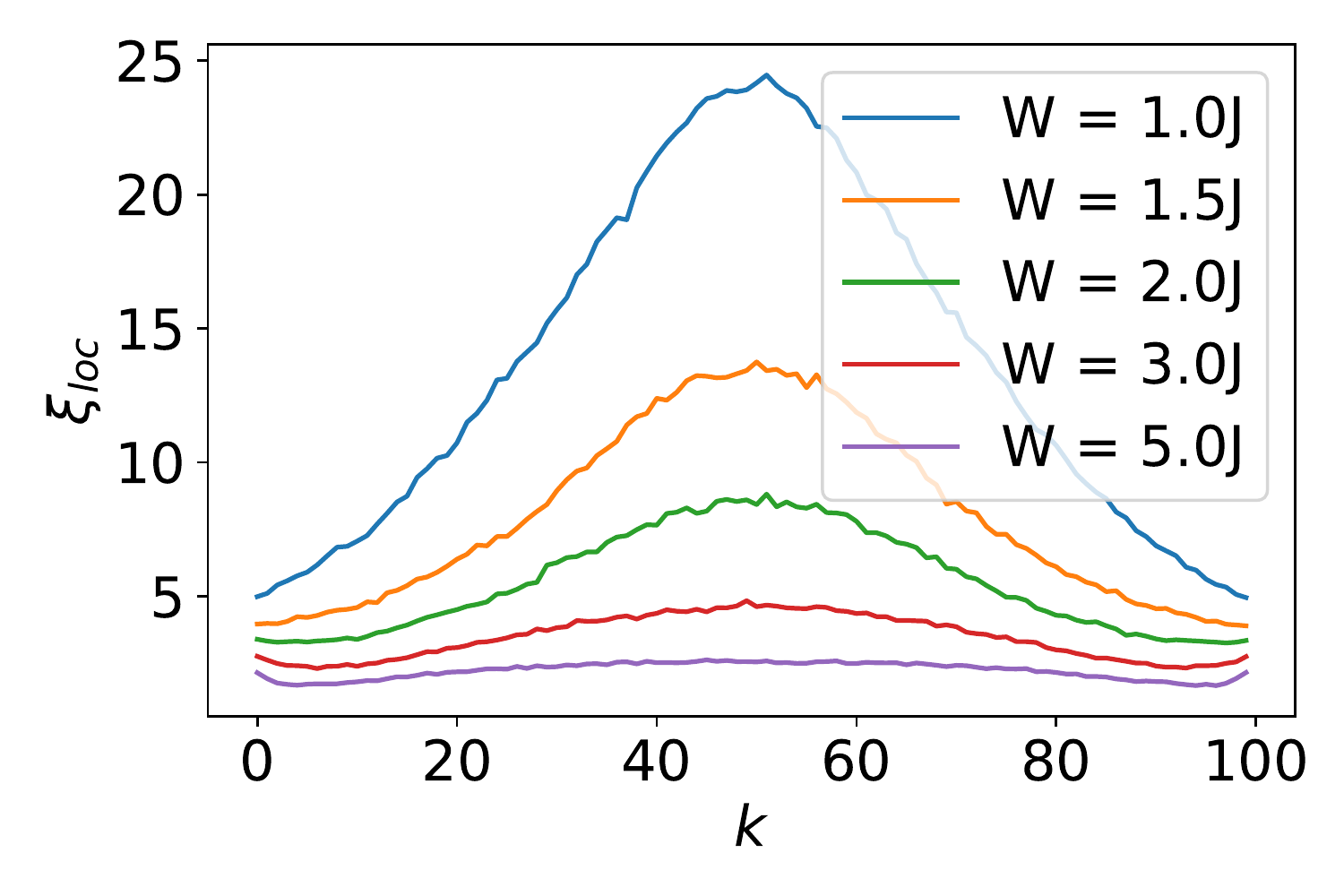}}
    \subfloat[]{\includegraphics[width=0.33\columnwidth]{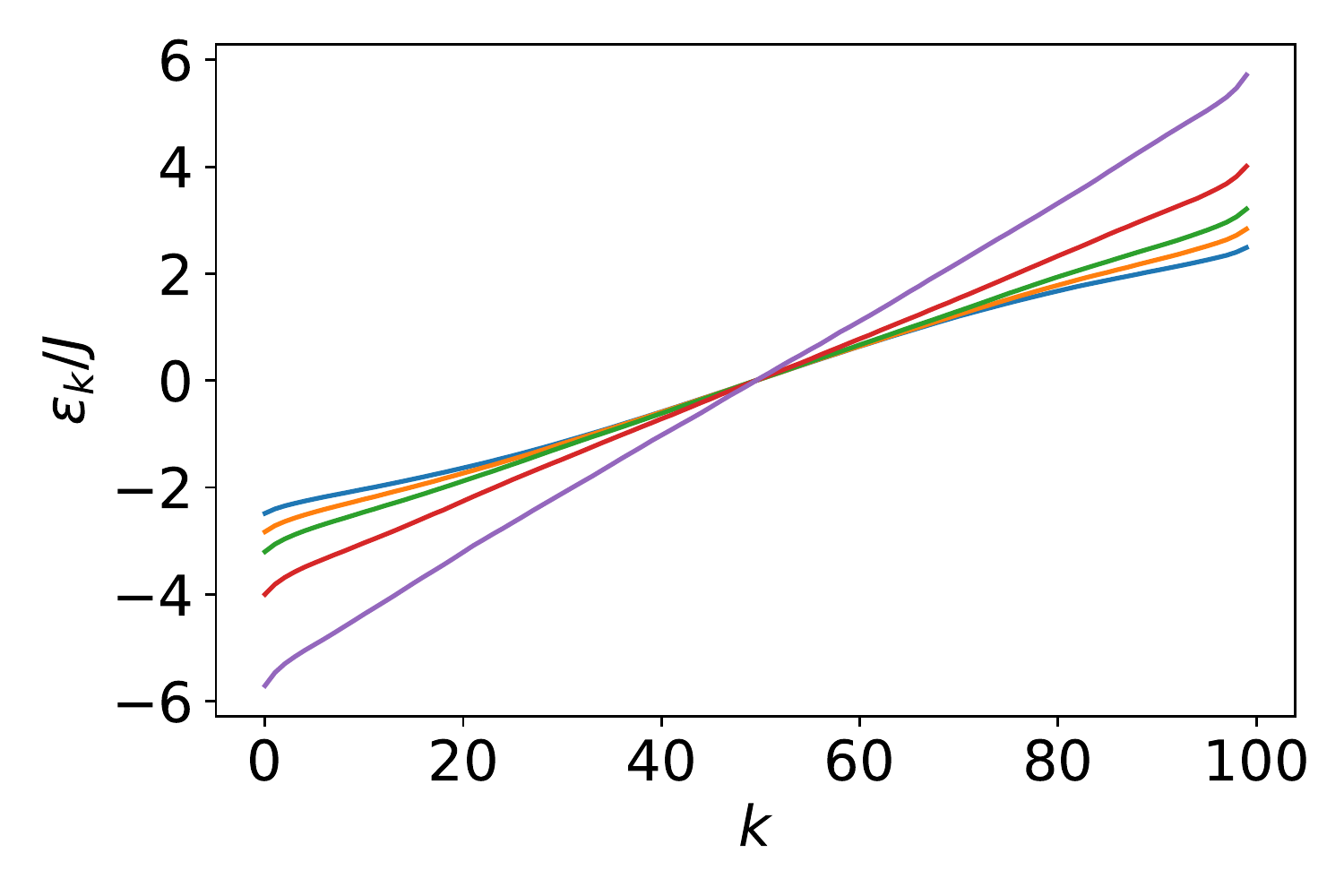}}
    \subfloat[]{\includegraphics[width=0.33\columnwidth]{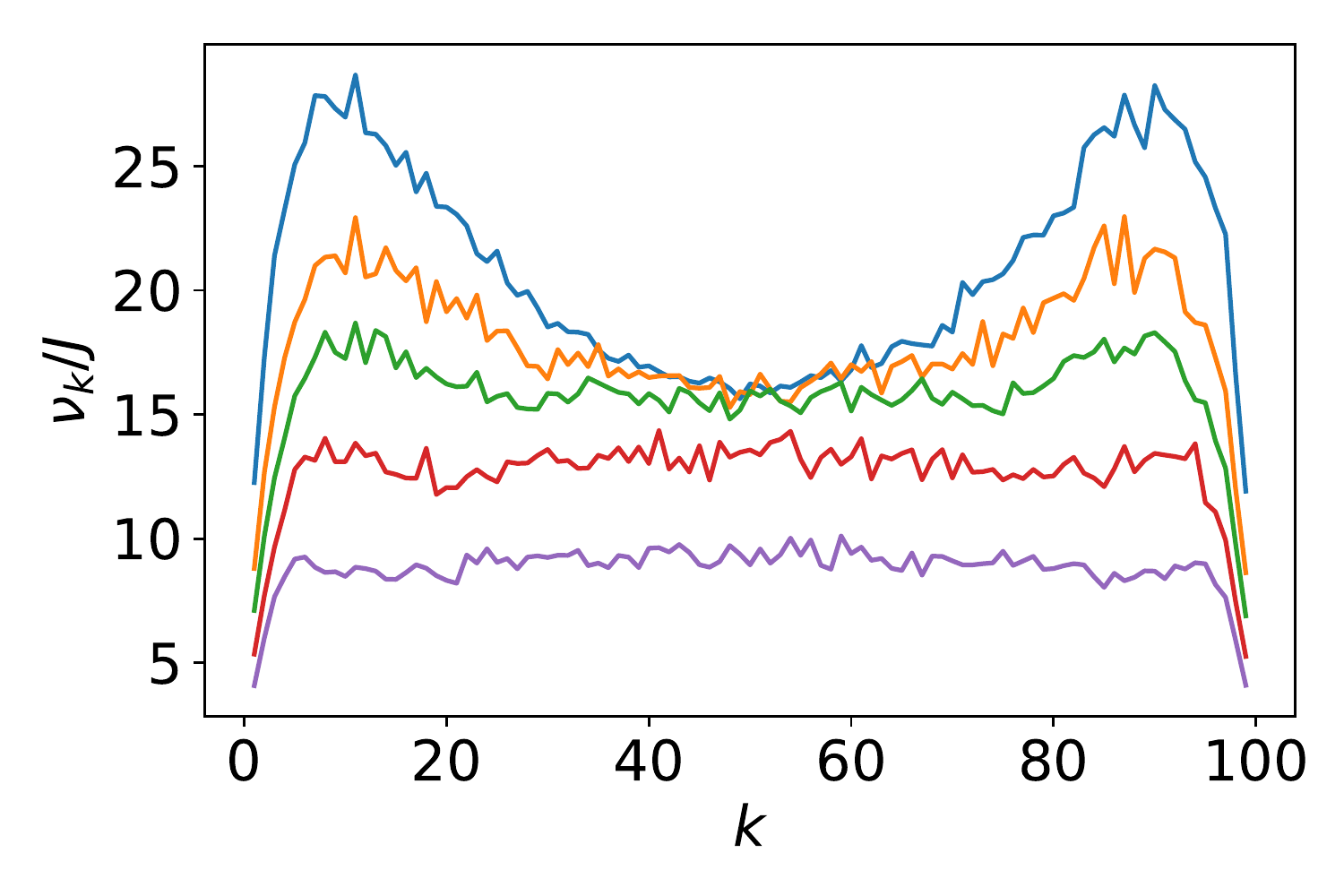}}
    \caption{(a)-(b) show some selected eigenstates for the Hamiltonian (\ref{H0}) at different disorder strengths. (c) and (d) show the dependence of localization length $\xi_{\text{loc}} \equiv 1/\sum_i{|\psi_k{(i)}|^4}$ of the eigenstates  $\psi_k(i)$ on the disorder strength $W$ and the eigenstate label $k$. (e) shows the eigenenergies $\varepsilon_k$ for some values of $W$. (f) shows the density of state $\nu_k \equiv 1/(\varepsilon_k-\varepsilon_{k-1})$. System size is $M = 100$.
    }\label{wave}
\end{figure*}
In this section we show some further data concerning the non-interacting system.
A straightforward basis transformation recasts ${\hat H}_0$ into the diagonalized form ${{\hat H}_0} = \sum_k {{\varepsilon _k}{\hat{n}_k}}$
with $\hat{n}_k = \hat{c}_k^\dag \hat{c}_k$ and $\hat{c}_k^\dag  = \sum_i {{\psi _k}\left( i \right)\hat{a}_i^\dag } $,
which creates a particle in the single-particle eigenstate $|k\rangle = \sum_i {{\psi _k}\left( i \right) |i\rangle } $
with energy $\varepsilon_k$ that are distributed  between $-2J-W$ and $2J+W$, as shown in Fig.~\ref{wave}~(e).

Figure~\ref{wave}(a), (b) show some selected eigenstates  for two different disorder strengths.
It is clear that as  disorder becomes stronger, the wavefunctions become more localized.
To characterize the localization of the wavefunction, we define the  localization length using the
inverse participation ratio $\xi_{\text{loc}} \equiv 1/\sum_i{|\psi_k{(i)}|^4}$.
As shown in Fig.~\ref{wave}(c), $\xi_{\text{loc}}$ decays rapidly as disorder increases.
Moreover, it also depends on the eigenenergy, as shown in Fig.~\ref{wave}(d), and $\xi_{\text{loc}}$ has its maximum in the center of the energy-band [Fig.~\ref{wave}(e)].
The density of states $\nu_k \equiv 1/(\varepsilon_k-\varepsilon_{k-1})$ (inverse of the energy gap between neighboring eigenstates)
also depends on the disorder strength $W$. From Fig.~\ref{wave}(f), it is clear that
larger values of $W$ lead to a smaller density of states.

\subsection{Comparison of the current from kinetic theory and quantum Monte-Carlo simulation}\label{comparison}

Figure \ref{I_dI_W}(a) shows the  current as a function of disorder strength $W$. The kinetic theory results (solid lines) agree well with the results obtained from quantum-jump Monte-Carlo simulation (markers). The agreement is better for lower temperature and weaker disorder. The reason is that the error of the Monte-Carlo simulation scales as $\Delta I/\sqrt{N_{MC}}$, with $N_{MC}$ being the number of trajectories for the Monte-Carlo simulation and $\Delta I$  the fluctuation of the current.  From (b) we can see that, the  fluctuation of the current $\Delta I$ increases with disorder strength $W$and temperature $T$. In addition, as shown in (a), the mean value of the current decreases with increasing $W$. These imply that to maintain a small relative error, much more trajectories are needed for stronger disorder and higher temperature. In other words, for a given number of trajectories, which is $1000$ in our numerical calculation, the error of Monte-Carlo simulation will be larger for stronger disorder and higher temperature.

\begin{figure*}[!htp]
    \centering
\subfloat[][ ]{\includegraphics[width=0.5\columnwidth]{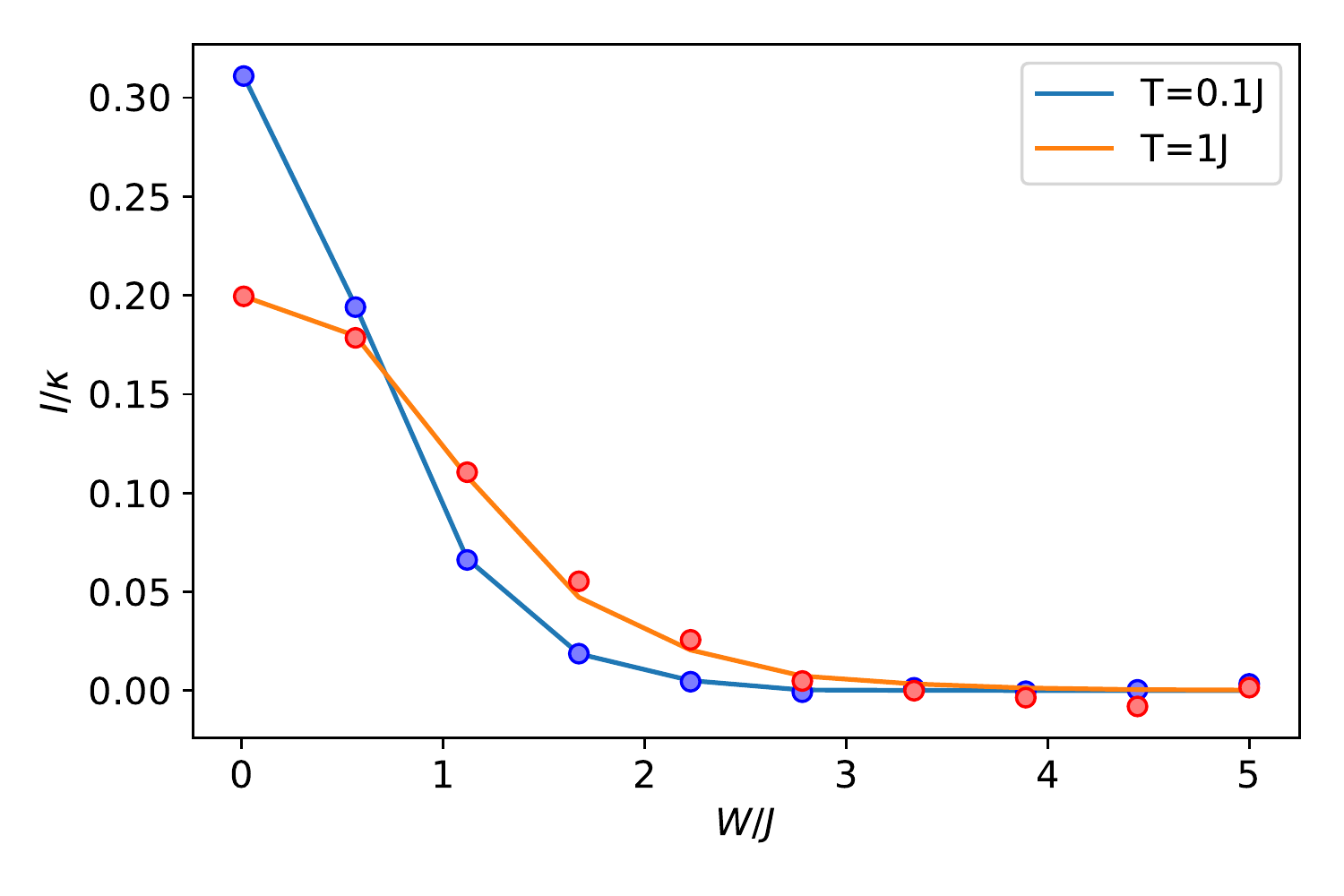}}
     \subfloat[][ ]{\includegraphics[width=0.5\columnwidth]{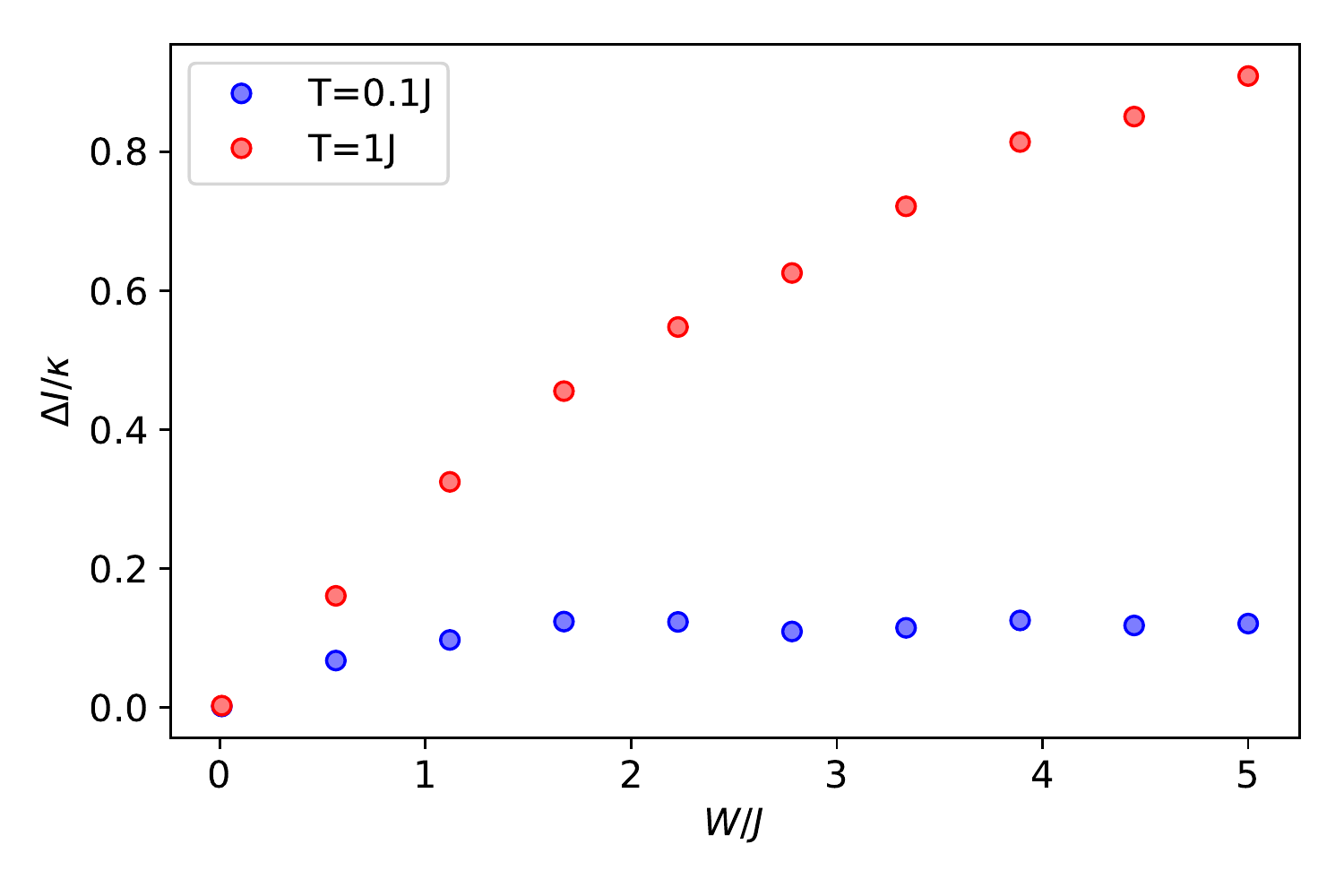}}
    \caption{ The current (a) and its fluctuation (b) as a function of the disorder strength $W$. The solid lines are the results of kinetic theory. The makers denote the results of quantum Monte-Carlo simulation. The results are averaged over $20$ disorder realizations.
    For a given disorder configuration, the  fluctuation of the current is given by $(\Delta I)_{j} = \sqrt{\langle I^2 \rangle_j - \langle I\rangle_j^2}$, where the average $\langle \cdot \rangle$ is taken over  $1000$ trajectories of Monte-Carlo simulations.
    The parameters are:  system size $M=50$, dissipation rate $\kappa =\gamma^2$,  interaction strength $V = 0$,  chemical potential imbalance $\delta \mu = J$.
    }\label{I_dI_W}
\end{figure*}

\newpage

\section*{References}

\bibliography{ref}{}
\bibliographystyle{iopart-num}

\end{document}